\RequirePackage{fix-cm}
\documentclass[smallextended]{svjour3}       
\usepackage{natbib}

\smartqed  
\usepackage{graphicx}
\journalname{Empirical Software Engineering}
\usepackage{amsmath,amssymb,amsfonts}
\usepackage{algorithmic}
\usepackage{textcomp}
\usepackage{rotating}
\usepackage{multirow}
\usepackage{booktabs}
\usepackage{tabularx}
\usepackage{mdframed}
\usepackage{longtable}

\usepackage{listings}
\usepackage{xcolor}
\usepackage{framed}
\usepackage{caption}
\usepackage{subcaption}
\usepackage{lscape}

\usepackage{pdfpages}

\usepackage{tikz}
\usetikzlibrary{shapes,arrows,positioning,decorations.pathreplacing}

\newcommand\YAMLcolonstyle{\color{red}\mdseries}
\newcommand\YAMLkeystyle{\color{black}\bfseries}
\newcommand\YAMLvaluestyle{\color{blue}\mdseries}

\makeatletter

\newcommand\language@yaml{yaml}

\expandafter\expandafter\expandafter\lstdefinelanguage
\expandafter{\language@yaml}
{
  captionpos=b,
  frame=single,
  keywords={true,false,null,y,n},
  keywordstyle=\color{darkgray}\bfseries,
  basicstyle=\YAMLkeystyle,                                 
  sensitive=false,
  comment=[l]{\#},
  morecomment=[s]{/*}{*/},
  commentstyle=\color{purple}\ttfamily,
  stringstyle=\YAMLvaluestyle\ttfamily,
  moredelim=[l][\color{orange}]{\&},
  moredelim=[l][\color{magenta}]{*},
  moredelim=**[il][\YAMLcolonstyle{:}\YAMLvaluestyle]{:},   
  morestring=[b]',
  morestring=[b]",
  literate =    {---}{{\ProcessThreeDashes}}3
                {>}{{\textcolor{red}\textgreater}}1     
                {|}{{\textcolor{red}\textbar}}1 
                {\ -\ }{{\mdseries\ -\ }}3,
}

\usepackage{hyperref}

\begin{document}


\newcommand{\atoml}{\texttt{atoml}}
\newcommand{\rev}[1]{#1}

\newcommand\setrow[1]{\gdef\rowmac{#1}#1\ignorespaces}
\newcommand\clearrow{\global\let\rowmac\relax}
\clearrow

\title{Exploring the relationship between performance metrics and cost saving potential of defect prediction models}
\titlerunning{Relationship between performance metrics and cost saving potential}

\author{
Steffen Tunkel \and
Steffen Herbold}

\institute{
Steffen Tunkel\\Institute of Computer Science, University of Goettingen, Germany\\
\email{steffen.tunkel@stud.uni-goettingen.de}
\vspace{5pt}\\
Steffen Herbold (corresponding author)\\Institute of Software and Systems Engineering, TU Clausthal, Germany\\
\email{steffen.herbold@tu-clausthal.de}
}

\date{Received: date / Accepted: date}

\maketitle

\begin{abstract}
\textit{Context:} Performance metrics are a core component of the evaluation of any machine learning model and used to compare models and estimate their usefulness. Recent work started to question the validity of many performance metrics for this purpose in the context of software defect prediction.

\textit{Objective:} Within this study, we explore the relationship between performance metrics and the cost saving potential of defect prediction models. We study whether performance metrics are suitable proxies to evaluate the cost saving capabilities and derive a theory for the relationship between performance metrics and cost saving potential.

\textit{Methods:} We measure performance metrics and cost saving potential in defect prediction experiments. We use a multinomial logit model, decision, and random forest to model the relationship between the metrics and the cost savings.

\textit{Results:} We could not find a stable relationship between cost savings and performance metrics. We attribute the lack of the relationship to the inability of performance metrics to account for the property that a small proportion of very large software artifacts are the main driver of the costs.

fact that performance metrics are incapable of accurately considering the costs associate with individual artifacts, which is required due to the exponential distribution of artifact sizes.

\textit{Conclusion:} Any defect prediction study interested in finding the best prediction model, must consider cost savings directly, because no reasonable claims regarding the economic benefits of defect prediction can be made otherwise. 

\end{abstract}

\section{Introduction}
\label{sec:introduction}

Software defect prediction is a popular field of research~\citep{Hall2011, Hosseini2017}. Regardless of the type of defect prediction (e.g., just-in-time~\citep{Kamei2013}, cross-project~\citep{Zimmermann2009}, cross-version~\citep{Amasaki2020}, heterogeneous~\citep{Nam2018}, or unsupervised~\citep{Nam2015}), the quality of defect prediction models is always evaluated by comparing predictions with labeled data. Consequently, the choice of suitable metrics for the evaluation is crucial for an accurate assessment of defect prediction models. 

Recently, researchers started to question the performance metrics that defect prediction researchers use to evaluate results in case studies~\citep{Herbold2019, Morasca2020, Yao2021}. \cite{Herbold2019} defined a model for the costs of defect prediction through the lens of quality assurance, missed defects, and costs for introducing and running a defect prediction model. Through his work, he derived necessary boundary conditions that must be fulfilled for defect prediction to save costs and showed that while metrics like \textit{precision} and \textit{recall} are correlated to costs, their absolute values are not good estimators for the cost saving potential of a defect prediction model. \cite{Morasca2020} consider the suitability of the Receiver Operating Characteristic (ROC) and the metric Area under the Curve (\textit{AUC}) for the evaluation of models. They argue that the family of classifiers that is evaluated through \textit{AUC} leads to misleading results, as it is unclear if the prediction models outperform trivial models. Instead, they argue, there is only a certain region of interest that should be evaluated, as other values could never be cost efficient. \cite{Yao2021} use a perspective that is not directly related to costs, but rather to the question if performance metrics are suitable to identify models that are better than randomness. They show that the commonly used criterion \textit{F-measure} is unreliable, as large values may not be related to good predictions at all, due to properties of the data. Instead, they show that Matthews Correlation Coefficient (\textit{MCC}) is a reliable estimator to determine if models are better than randomness. However, the relationship between \textit{MCC} and costs is unclear. There are also earlier results, e.g., by \cite{Rahman2012} that indicate problems with performance metrics if cost related aspects are ignored, e.g., the size of predicted artifacts. 

Thus, while all these studies consider different angles, the conclusions are similar: performance metrics are often problematic because their absolute values are unreliable proxies for the practical relevance of the prediction performance and the relation between performance metrics and cost savings is rather indirect. This is perhaps also the reason why there is no consensus in the defect prediction community regarding the question which metrics should be used, which leads to a large number of different criteria that are used~\citep{Hosseini2017, Herbold2017} and has potentially severe effects on the validity of our work like researcher bias~\citep{Shepperd2014, tantithamthavorn2016comments, Shepperd2018}. 

Within this study, we shed light on the relationship between performance metrics and their relationship to cost savings by studying the following research question:

\begin{itemize}
\item \textbf{RQ} Which performance metrics are good indicators for cost saving potential of defect prediction models?
\end{itemize}

We study the indirect relationship between performance criteria and cost saving potential empirically with the goal to foster an understanding of this relationship. We derive four levels of cost saving potential from Herbold's cost model~\citep{Herbold2019} and try to model this cost saving potential as dependent variable through other metrics as independent variables. We introduce several confounding variables, which we believe could also be good estimators for the cost saving potential of defect prediction models, even though they are not measuring performance. Because our study design has many options (defect prediction model selection, independent variables, confounding variables, models to establish relationships between variables), we pre-registered the protocol of our study~\citep{herbold2021exploring}. The main contributions of our study are the following. 

\begin{itemize}
    \item We did not find a generalizable relationship between performance metrics and the cost saving potential of software defect prediction through our empirical study. We provide a mathematical explanation by considering the formulation of the costs and the metrics. The analysis revealed that the likely reason for the lack of a relationship is the small proportion of very large software artifacts as the main driver of the costs because this means that a small proportion of the data drives most of the costs. 
    \item We suggest that future research always considers costs directly, if the economic performance of defect prediction models is relevant. This means that all studies aiming to find the ``best'' defect prediction model, should consider this criterion. Otherwise, no claims regarding being better in a use case that involves the prediction of defects in a company to guide quality assurance should be made. 
    \item We find that the chance that release-level defect prediction models are at all cost saving is mediocre and was for the best model we observed only at 63\%. This means that for more than one third of the cases, it can never make economic sense to use release level defect prediction. We note that while this finding is restricted to the approaches we use, we obtained this based on good models from prior benchmark studies.
\end{itemize}

The remainder of this article is structured as follows. In Section~\ref{sec:related-work}, we discuss other studies on the cost of defect prediction. Then, we introduce our pre-registered research protocol in Section~\ref{sec:protocol} and present the results of the execution of this research protocol in Section~\ref{sec:results}. We discuss the results to answer our research question in Section~\ref{sec:discussion}, including the threats to the validity of our work. Finally, we conclude in Section~\ref{sec:conclusion}. 

\section{Related Work}
\label{sec:related-work}

Within this section, we discuss work related to the relationship between costs and defect prediction. We reference many different performance metrics within this discussion. The definitions of these metrics can be found in Section~\ref{sec:variables}. 

To the best of our knowledge, there is no work that tries to empirically establish the relationship between different performance metrics and cost saving potential of defect prediction models. The closest relationship to our work is the analysis of the costs of defect prediction by \cite{Herbold2019}, who introduced a cost model for defect prediction, as well as mandatory boundary conditions that must be fulfilled to outperform not using defect prediction. Not using defect prediction can be considered as trivial defect prediction models, that either predict everything as defective or everything is non-defective. There are two aspects relevant to our work: first, we use the cost model proposed to define our dependent variable, which we discuss in Section~\ref{sec:dependent}. Moreover, as part of the validation of the introduced cost model, aspects of the model are directly compared to \textit{recall} and \textit{precision}. The analysis found that there are two types of projects, regarding cost saving potential: those, where a high \textit{recall} is sufficient, even if the \textit{precision} is low ($<$25\%), and those with a high \textit{recall}, that require a mediocre \textit{precision} ($\geq$25\% and $<$50\%). Notably, \cite{Herbold2019} found that none of the studied projects required a high \textit{precision} to allow cost savings. The results further show that there are no linear relationships between cost savings and \textit{recall} and \textit{precision}, meaning that a 10\% increase in a confusion matrix-based metric would not necessarily translate to a 10\% increase in cost savings, where the relative change could be both lower or higher. Overall, \cite{Herbold2019} found that thresholds for confusion matrix-based metrics are not suitable as criteria for success of defect prediction in terms of cost savings. 

While the work by \cite{Herbold2019} is a good starting point for the analysis of the relationship between performance metrics and costs, including initial results, the study has several weaknesses that we try to resolve within this article. First, no actual defect predictions were conducted by \cite{Herbold2019}. Instead, only simulations were carried out to analyze trends. Moreover, these simulations only considered \textit{recall} and \textit{precision}. In contrast, we collect data about defect prediction on a large scale. We also consider a set of 20 performance metrics, including criteria that go beyond the confusion matrix, as well as ten confounding variables to look for explanations of cost savings in general properties of projects, independent of the defect prediction model. 

\cite{Morasca2020} do not directly consider costs, but rather explore how the \textit{AUC} is unsuitable for the evaluation of defect prediction models. Their criticism of \textit{AUC} is that the complete area under the ROC curve is considered. However, they demonstrate that this can lead to misleading results because sometimes trivial models have better values than the defect prediction model. They argue that the region that is considered for the \textit{AUC} consideration should be restricted to performance values, where the defect prediction model could actually beat trivial models. As trivial models, the authors consider random models in addition to predicting everything defective or non-defective. Though the work by \cite{Morasca2020} does not consider costs, it shows that performance metrics can be misleading, as they often hide that trivial approaches would perform better. Since \cite{Herbold2019} uses the same boundaries, it follows that the performance measured of \textit{AUC} is also, at least to some degree, decoupled from costs. Within this work, we fill in the gaps between the results established by \cite{Morasca2020} and costs. We study the proposed variant of the \textit{AUC} that only takes regions better than trivial models into account as an independent variable. This allows us to gain insights into the question if it is sufficient to measure confusion matrix-based metrics to determine if defect prediction models are better than trivial models, without taking other cost factors, like the size of software artifacts, into account.

\cite{Yao2021} also do not study costs, but rather the suitability of performance metrics to detect if defect prediction models perform better than a random model. They show that especially the often used \textit{F-measure} is problematic. Research has shown that the \textit{F-measure} is unreliable in the presence of class level imbalance and that other approaches are superior, e.g., the correlation based approach of \textit{MCC} \citep{Luque2019}. \cite{Yao2021} confirm that these problems are also relevant for defect prediction: about one fifth of conclusions regarding performance differences based on the \textit{F-measure} are wrong or misleading. However, while \cite{Yao2021} make a compelling case that \textit{MCC} should be used for the evaluation of defect prediction studies, this is only true for general prediction performance. Within our study, we determine if \textit{MCC} is also well suited as a proxy for measuring cost effectiveness. 

There is also prior work that proposes to use cost for the evaluation of defect prediction models. For example, \cite{Khoshgoftaar1998} propose to treat the costs of false positives (wrongly predicting defects) and false negatives (missing defects) differently. They argue that this allows the definition of the expected costs of misclassifications. However, since this is only a weighting of different entries in a confusion matrix, other aspects regarding costs, especially difference in costs due to the size of software artifacts, are ignored. Similar approaches are used by \cite{Drummond2006} for the definition of cost curves, or by \cite{Liu2010} with a normalization of the costs. 

There are also several performance metrics that try to measure costs by considering the size of software artifacts. \cite{Ohlsson1996}, \cite{Arisholm2006}, \cite{Rahman2012}, and \cite{Hemmati2015} all used similar approaches: they define variants of ROC curves, where one axis of the ROC curves contains a size-based metric, e.g., the lines of code of software artifacts predicted as defective. In comparison to confusion matrix-based metrics, these metrics take the difference in costs due to artifact sizes into account. However, as \cite{Morasca2020} showed, ROC based metrics may be misleading because they ignore the performance that could be achieved with trivial models. Within our work, we bridge these two aspects, by using the size-based performance metrics as independent variables of our study to establish their relationship with cost saving potential in relationship to trivial models.

\section{Research Protocol}
\label{sec:protocol}

Within this section, we discuss the research protocol, i.e., the research strategy we pre-registered~\citep{herbold2021exploring} for our exploratory study. The description and section structure of our research protocol is closely aligned with the pre-registration. All deviations are described as part of this section and summarized in Section~\ref{sec:deviations}.

\subsection{Notation}

We use the following notation for the specification of variables. 

\begin{itemize}
  \item $S$ is the set of software artifacts for which defects are predicted. Examples for software artifacts are files, classes, methods, or changes to any of the aforementioned.
  \item $h: S \to \{0, 1\}$ is the defect prediction model, where $h(s)=1$ means that the model predicts a defect in an artifact $s \in S$. Alternatively, we sometimes use the notation $h'(s) \to [0,1]$ with a threshold $t$ such that $h(s)=1$ if and only if $h'(s) > t$, in case performance metrics require scores for instances.
  \item $D$ is the set of defects $d \subseteq S$. Thus, a defect is defined by the set of software artifacts that are affected by the defect.
  \item $D_\textit{PRED} = \{d \in D: \forall s \in d~|~h(s) = 1\}$ is the set of predicted defects. Hence, a defect is only predicted successfully, if the defect prediction model predicts all files affected by the defect. Through this, we account for the $n$-to-$m$ relationship between artifacts and defects, i.e., one file may be affected by multiple bugs and one bug may affect multiple files.
  \item $D_\textit{MISS} = \{d \in D: \exists s \in d~|~h(s)=0\} = D \setminus D_\textit{PRED}$  is the set of missed defects.
  \item $S_\textit{DEF} = \{s \in S: \exists d \in D~|~s \in d\}$ is the set of software artifacts that are defective.
  \item $S_\textit{CLEAN} = \{s \in S: \nexists d \in D~|~s \in d\}$ is the set of software artifacts that are clean, i.e., not defective.
  \item $h^*: S \to \{0,1\}$ is the target model, where each prediction is correct, i.e., $h^*(s)= \begin{cases}1 & s \in S_\textit{DEF} \\ 0 & s \in S_\textit{CLEAN}\end{cases}$
  \item $tp = |\{s \in S: h^*(s)=1~\text{and}~h(s)=1\}|$ are the artifacts that are affected by any defect and correctly predicted as defective. 
  \item $fn = |\{s \in S: h^*(s)=1~\text{and}~h(s)=0\}|$ are the artifacts that are affected by any defect and are missed by the prediction model. 
  \item $tn = |\{s \in S: h^*(s)=0~\text{and}~h(s)=0\}|$ are the artifacts that are clean and correctly predicted as clean by the prediction model. 
  \item $fp = |\{s \in S: h^*(s)=0~\text{and}~h(s)=1\}|$ are the artifacts that are clean and wrongly predicted as defective by the prediction model.
  \item $size(s)$ is the size of the software artifact $s \in S$. Within this study, we use $\textit{size}(s) = \textit{LLOC}(s)$, where \textit{LLOC} are the logical lines of code of the artifact.\footnote{Number of non-empty and non-comment code lines.} 
  \item $C = \frac{C_\textit{DEF}}{C_\textit{QA}}$ is the ratio between the expected costs of a post release defect $C_\textit{DEF}$ and the expected costs for quality assurance per size unit $C_\textit{QA}$. Costs for quality assurance for an artifact $qa(s)$ are, therefore, $qa(s) = size(s) \cdot C_{QA}$.
\end{itemize}

\subsection{Research Question}
\label{sec:rq}

The goal of our study is to address the following research question. 

\begin{itemize}
\item \textbf{RQ:} Which performance metrics are good indicators for cost saving potential of defect prediction models?
\end{itemize}

Because the literature does not provide suitable evidence to derive concrete hypotheses, we conduct an exploratory study that generates empirical evidence which we use to derive a theory that answers this question. While the prior work provides an analytic mathematical analysis of costs and describes why at least some performance criteria should be unsuitable for the estimation of the cost saving potential,\footnote{\textit{precision} and \textit{recall} by \cite{Herbold2019}, \textit{AUC} by \cite{Morasca2020}, and \textit{F-measure} by \cite{Yao2021}} the set of performance criteria considered so far is incomplete and the interactions between performance criteria were not yet considered. Especially the unresolved question if combinations of different performance criteria could be a suitable proxy for cost estimation is yet unexplored, such that we cannot derive hypotheses from prior data to drive a confirmatory study.

Our investigation of this research question is driven by empirical data that we collect through an experiment. The alternative would be to study the problem analytically, e.g., by analyzing the mathematical properties of the performance criteria in comparison to the cost models. However, this alternative would be limited, as we would need a mathematical model for the size of defective and clean software artifacts as key drivers of the costs. Moreover, the study of interactions between multiple performance criteria could become very complicated because many different performance criteria would need to be considered at once. Through our empirical approach, we have accurate data regarding the size of artifacts and can use different models we infer from the data to understand the relationship between performance metrics and costs. This does not mean we abandon the analytic approach, but rather that we postpone such considerations for the theory building. Once we have established the relationships between our variables empirically, we try to find reasons for these relationships in the mathematical description of these metrics to establish causal relationships.

\subsection{Variables}
\label{sec:variables}

In our study we have one ordinal variable as dependent variable that we model through twenty independent variables and ten confounding variables. 

\subsubsection{Dependent Variable}
\label{sec:dependent}

The key question is how we define the cost saving potential of defect prediction. We use the cost model by \cite{Herbold2019} for this purpose. Herbold proved boundary conditions that must be fulfilled by defect prediction models if they should perform better than trivial models, i.e., either predicting all instances as defective or non-defective. Under the assumption that we can use the size of artifacts $\textit{size}(s)$ as proxy for the quality assurance effort and that the quality assurance always finds defects, we get

\begin{equation}
\textit{lower} = \frac{\sum_{s \in S: h(s)=1} \textit{size}(s)}{|D_\textit{PRED}|}
\end{equation}

as lower bound on the cost ratio $C$ and

\begin{equation}
\textit{upper} = \frac{\sum_{s \in S: h(s)=0} \textit{size}(s)}{|D_\textit{MISS}|}
\end{equation}

as the upper bound on the cost ratio. The lower bound is derived from the condition that a defect prediction model must be better than a trivial model that does not predict any defects and measures the effort for quality assurance to find correctly predicted defects in relation to the number of defects found. The upper bound is derived from the condition that a defect prediction model must be better than a trivial model that assumes everything is defective and measures the costs that are saved by not applying quality assurance to artifacts that are not predicted as defective in relation to the missed defects. Based on the bounds, the project specific ratio between the costs of a defect and the costs for quality assurance can be bounded. This value is not known in general and there are no estimations for this within the current body of research~\citep{Herbold2019}. Still, the difference between the boundaries is a good way to estimate the cost saving potential of defect prediction models. A larger difference between the boundaries means that the defect prediction saves cost for more cost ratios and, consequently, more projects. Moreover, the further away from the boundaries the actual value of $C$ is, the higher the cost savings, since the cost saving potential is minimal at the boundaries and increases with the distance from the boundary. Consequently, a larger difference between the boundaries does not only increase the likelihood that projects are able to save costs, but also means that the cost savings can be higher.

Thus, we base our dependent variable for the cost saving potential on the difference between the upper and lower bound. One option would be to model the dependent variable as a continuous variable

\begin{equation}
    \textit{diff} = \textit{upper}-\textit{lower}. 
\end{equation}

However, the mathematical properties of \textit{lower} and \textit{upper} make \textit{diff} an unsuitable choice as continuous dependent variable. The lower bound is undefined for defect prediction models that predict no defects because both numerator and denominator become zero. Vice versa, the upper bound is undefined for defect prediction models that predict everything as defective. This is a consequence of the construction of the boundaries, but no practical problem when it comes to evaluating the cost saving potential: a trivial model is not a real defect prediction model. Hence, the undefined value indicates that there is no cost-saving potential. Moreover, both boundaries may become infinite. The lower bound is $\infty$ if no defect is predicted correctly but at least one artifact is predicted as defective. In that case, \textit{diff} is $-\infty$. Vice versa, the upper bound is $\infty$ if no defect is missed and at least one artifact is predicted as clean. In that case, \textit{diff} is $\infty$. 

To avoid these numeric pitfalls, we use the ordinal variable \textit{potential} that we define using decadic logarithmic bins for \textit{diff}, such that
\begin{equation*}
\textit{potential} =
\begin{cases}
\textit{none} & \text{if}~\textit{diff}~\in (-\infty, 0]~\text{or}~\textit{diff}~\text{is}~\textit{NaN} \\
\textit{medium} & \text{if}~\textit{diff}~\in (0, 1000] \\
\textit{large} & \text{if}~\textit{diff}~\in (1000, 10000] \\
\textit{extra large} & \text{if}~\textit{diff}~\in (10000, \infty]. 
\end{cases}
\end{equation*}
where \textit{NaN} means that the value of \textit{diff} is undefined, as our dependent variable. The interpretation of \textit{potential} is best explained using an example. Let $\textit{lower}=500$ and $\textit{upper}=1550$. This means that the model saves costs if $C \in [500, 1550]$. Thus, the cost for defects must be greater than quality assurance costs for 500 LLOC and less than the quality assurance costs for 1550 LLOC. Consequently, we have a \textit{large} range of $\textit{diff}=1050$ where we can save costs. 

This definition of \textit{potential} deviates from our registration protocol, that included the additional levels \textit{negligible} for the interval $(0, 10]$ and \textit{small} for the interval $(10, 100]$. These levels were underrepresented in our data with less than 0.2\% of the overall data. Consequently, we merged these bins with the \textit{medium} bin, as this reduces the complexity of our analysis, without a negative effect on the sensitivity.

\subsubsection{Independent Variables}

The independent variables of our study are performance metrics of defect prediction experiments. The selection of these metrics is based on two systematic literature studies on defect prediction research~\citep{Hosseini2017, Herbold2017}, with the goal to cover a broad set of metrics that are currently used. Overall, we identified 20 performance metrics. For each of these metrics we cite one publication where the metric was used. 

\begin{enumerate}
  \item $\textit{recall} = pd = tpr = \textit{completeness} = \frac{tp}{tp+fn}$ measures the percentage of defective artifacts that are defected and was, e.g., used by \cite{Watanabe2008}.
  \item $\textit{precision} = \textit{correctness} = \frac{tp}{tp+fp}$ measures the percentage of artifacts that are predicted as defective that is correct and was, e.g., used by \cite{Watanabe2008}.
  \item $pf = \textit{fpr} = \frac{fp}{tn+fp}$ is the probability of a false alarm, also known as the false positive rate and was, e.g., used by \cite{Peters2013}.
  \item $\textit{F-measure} = 2 \cdot \frac{\textit{recall} \cdot \textit{precision}}{\textit{recall}+\textit{precision}}$ is the harmonic mean of \textit{precision} and \textit{recall} and was, e.g., used by \cite{Kawata2015}.
  \item $\textit{G-measure} = 2 \cdot \frac{\textit{recall} \cdot (1-pf)}{\textit{recall}+(1-pf)}$ is the harmonic mean of \textit{recall} and the opposite probability of a false alarm and was, e.g., used by \cite{Peters2013}. 
  \item $\textit{balance} = 1-\frac{\sqrt{(1-\textit{recall})^2+pf^2}}{\sqrt{2}}$ is the normalized euclidean distance between the observed performance and optimal performance as depicted by a ROC curve \cite{Turhan2013}.
  \item $\textit{accuracy} = \frac{tp+tn}{tp+fp+tn+fn}$ is the percentage of artifacts that are predicted correctly and was, e.g., used by \cite{Zimmermann2009}. 
  \item $\textit{error} = \frac{fp+fn}{tp+fp+tn+fn}$ is the percentage of artifacts that are predicted incorrectly and was, e.g., used by \cite{Liu2010}. 
  \item $\textit{error}_\textit{TypeI} = \frac{fp}{tp+fn}$ is the Type I error rate, i.e. the ratio of artifacts that are incorrectly predicted as defective and the actually defective artifacts and was, e.g., used by \cite{Liu2010}.
  \item $\textit{error}_\textit{TypeII} = \frac{fn}{tn+fp}$ is the Type II error rate, i.e. the ratio of artifacts that are incorrectly predicted as clean and the actually clean artifacts and was, e.g., used by \cite{Liu2010}.
  \item $\textit{MCC} = \frac{tp \cdot tn - fp \cdot fn}{\sqrt{(tp+fp)(tp+fn)(tn+fp)(tn+fn))}}$ is the $\chi^2$ correlation between the predictions and the actual values and was, e.g., used by \cite{Zhang2015}. 
  \item $\textit{consistency} = \frac{tp\cdot(tp+fp+tn+fn)-(tp+fn)^2}{(tp+fn)\cdot(tn+fp)}$ is a metric for model stability and was, e.g., used by \cite{He2015}.
\end{enumerate}

Three variants of the Area Under the Curve (\textit{AUC}) are used. \textit{AUC} is distributed between zero and one. The variants of \textit{AUC} are defined using different variants of a Receiver Operating Characteristic (ROC). The shape of the ROC curve is determined by evaluating $h'$ for all possible thresholds of $t$. Consequently, the ROC based metrics are called threshold independent. 
\begin{enumerate}
  \setcounter{enumi}{12}
  \item \textit{AUC} uses the \textit{pf} versus \textit{recall} as ROC and was, e.g., used by \cite{Zhang2015}.
  \item $\textit{AUC}_\textit{Alberg}$ uses the \textit{recall} and the percentage of modules considered to define the ROC \citep{Ohlsson1996} and was, e.g., used by \cite{Rahman2012}.
  \item $\textit{AUC}_\textit{recall, pf}$ where only the region between the ROC curve is considered, where the values are better than the \textit{recall} and $pf$ achieved by a random model that knows the class level imbalance, i.e., the ratio between defective and non-defective artifacts. This metric is taken from the analysis of problems with \textit{AUC} for defect prediction by \cite{Morasca2020}.
\end{enumerate}

Furthermore, measures related to the cost required for reviewing effort are used.
\begin{enumerate}
  \setcounter{enumi}{15}
    \item $\textit{NECM}_{10} = \frac{fp + 10\cdot fn}{tp+fp+tn+fn}$ as the normalized expected cost of misclassification, where 10 is the ratio between the costs of a false negative (Type II error) and the costs of a false positive (Type I error). Hence, this metric assume that missing defects is ten times as costly as additional quality assurance. This metric was, e.g., used by \cite{Liu2010}. 
  \item $\textit{NECM}_{25} = \frac{fp + 25\cdot fn}{tp+fp+tn+fn}$ is the same as $\textit{NECM}_{10}$ but with a ratio of 25 and was, e.g., used by \cite{Liu2010}.
  \item $\textit{cost} = \sum_{s \in S} h(s)\cdot size(s)$ is the cost for quality assurance for all predicted artifacts and was, e.g, used by \cite{Canfora2013}.
  \item $\textit{NofB}_{20\%}$, i.e., the number of bugs found when inspecting 20\% of the code and was, e.g., used by \cite{Zhang2015a}. 
  \item $\textit{NofC}_{80\%}$, i.e., number of classes visited until 80\% of the bugs are found and was, e.g., used by \cite{Jureczko2010}.
\end{enumerate}

\subsubsection{Confounding Variables}

We use several confounding variables that are not related to the performance, but rather to properties of the training and test data sets and the relationship between the two data sets.  

\begin{enumerate}
  \item $\textit{bias}_\textit{train}$, i.e., the percentage of instances in the training data that are defective.
  \item $\textit{bias}'_\textit{train}$, i.e., the percentage of instances in the training data that are defective after pre-processing, e.g., after oversampling of the minority class. 
  \item $\textit{bias}_\textit{test}$, i.e., the percentage of instances in the test data that are defective. 
  \item $\Delta_\textit{ratio} \textit{bias} = \frac{\textit{bias}_\textit{test}}{\textit{bias}_\textit{train}}$, i.e., the ratio of the bias of the training and the test data.
  \item $\Delta_\textit{ratio} \textit{bias}' = \frac{bias_{test}}{bias'_{train}}$, where $\textit{bias}'_\textit{train}$ is the bias of the training data after pre-processing, e.g., after oversampling of the minority class.
  \item $\textit{prop}_\textit{def}^{1\%} = \frac{\sum_{s \in LD} \textit{size}(s)}{\sum_{s \in S_\textit{DEF}}}$, where $LD \subset S_\textit{DEF}$ are the 1\% largest defective instances in the test data.
  \item $\textit{prop}_\textit{clean}^{1\%} = \frac{\sum_{s \in LC} \textit{size}(s)}{\sum_{s \in S_\textit{CLEAN}}}$, where $LC \subset S_\textit{CLEAN}$ are the 1\% largest clean instances in the test data.
  \item $N_\textit{train}$, i.e., the number of instances in the training data.
  \item $N'_\textit{train}$, i.e., the number of instances in the training data after pre-processing. 
  \item $N_\textit{test} = tp+fp+tn+fn$, i.e., the size of the test data set.
\end{enumerate}

The first five confounding variables are selected based on the rationale that the ratio of defects may have a strong influence on the results. Finding defects is a lot harder if only 5\% of instances are defective in comparison to 50\% of data being defective. This should have an influence of the performance metrics and also the ability to save costs. However, this would not be a property of a good prediction model, but rather a project characteristic. Hence, it is important to determine if the bias has a strong influence on the cost saving potential, that possibly even dominates the prediction performance.

The confounding variables $\textit{prop}_\textit{def}^{1\%}$ and $\textit{prop}_\textit{clean}^{1\%}$ are based on the idea of \textit{super instances}, i.e., few very large instances who have an oversized impact on the costs. If larger values of these variables are useful to model the cost saving potential, this would indicate that the existence of such super instances is relevant for the cost saving potential. Interestingly, the impact on the costs of super instances is different for the defective and clean case. For the largest defective instances it may actually be best if the model misses them because their quality assurance is expensive in comparison to predicting bugs in smaller instances. Thus, missing them could actually be good for the lower boundary of the cost saving range, which is driven by the quality assurance effort for finding correctly predicted bugs. For the largest clean instances, mispredicting them would mean that a large potential to save quality assurance costs would not be used, which would be bad for the upper boundary of the costs.

The other three confounding variables are selected because the amount of data is crucial for the training of machine learning models. In theory, more training data should lead to better models with less overfitting. Such models should be able to better save costs. More test data should reduce possibly random effects and lead to more stable results. Thus, small projects may randomly lead to a very good or very bad performance because the evaluation is not about finding a large population of defects, but rather only very few defects. 

\subsection{Datasets}
\label{sec:data}

We use defect prediction data by \cite{Herbold2022} that contains data for 398 releases of 38 Java projects. Each instance in a release represents a Java production file. For each bug, the data contains which instances are affected. This allows us to respect the $n$-to-$m$ relationship between the files and the bugs that is required for the calculation of the $lower$ and $upper$ cost boundaries. To the best of our knowledge, the data by \cite{Herbold2022} is the only public defect prediction data set that contains this information. While the full data set provides 4,198 metrics, we use only the 66 static product metrics, i.e., the static file level metrics and the summation of the class level metrics. The initial experiments with the data by \cite{Herbold2022} indicate that the drop in performance due to this reduced feature set is likely small. However, in terms of execution time when training machine learning models, the difference is huge. Since we are not interested in finding the best machine learning model for defect prediction, but rather try to find out how performance metrics are related to cost saving potential, the training of more classifiers is more important than minor improvements in prediction performance. 

\begin{table}
\caption{Apache projects and releases used for the empirical study. The table shows the number of releases per project with the number of releases with at least 100 files and at least five defective files in parentheses, the range of the number of files and the ratio of defective files (in \%) for the releases.}
\label{tbl:projects}
\centering
\begin{tabular}{lrcc}
\textbf{Project} & \textbf{\#Releases} & \textbf{\#Files} & \textbf{Defect Ratio (\%)} \\
\hline\hline
ant-ivy & 6 (6) & 240-474 & 2.9-14.2 \\
archiva & 7 (6) & 430-467 & 0.7-6.9 \\
calcite & 16 (16) & 1075-1415 & 3.3-16.6 \\
cayenne & 2 (2) & 1578-1708 & 1.2-2.6 \\
commons-bcel & 6 (3) & 325-378 & 0.0-3.5 \\
commons-beanutils & 10 (1) & 3-104 & 0.0-8.0 \\
commons-codec & 11 (0) & 14-64 & 0.0-12.5 \\
commons-collections & 9 (5) & 26-301 & 0.3-5.2 \\
commons-compress & 17 (12) & 61-201 & 3.0-27.9 \\
commons-configuration & 14 (7) & 29-240 & 2.9-28.6 \\
commons-dbcp & 11 (0) & 32-56 & 1.8-35.9 \\
commons-digester & 14 (3) & 14-157 & 0.0-6.4 \\
commons-io & 11 (5) & 34-115 & 3.5-22.2 \\
commons-jcs & 6 (3) & 213-368 & 0.5-4.4 \\
commons-jexl & 6 (0) & 52-85 & 0.0-9.6 \\
commons-lang & 16 (5) & 26-138 & 0.0-22.2 \\
commons-math & 13 (12) & 106-914 & 0.0-7.2 \\
commons-net & 15 (13) & 95-270 & 1.9-19.8 \\
commons-scxml & 5 (0) & 72-79 & 6.3-19.0 \\
commons-validator & 7 (0) & 17-63 & 0.0-31.8 \\
commons-vfs & 4 (2) & 236-262 & 0.8-11.0 \\
deltaspike & 16 (15) & 56-725 & 1.1-10.5 \\
eagle & 3 (1) & 682-1388 & 0.0-2.6 \\
giraph & 3 (2) & 105-753 & 0.0-6.3 \\
gora & 8 (5) & 97-210 & 0.0-11.3 \\
jspwiki & 12 (2) & 13-730 & 0.5-7.7 \\
knox & 13 (13) & 388-763 & 1.1-12.7 \\
kylin & 11 (11) & 379-1006 & 2.1-8.1 \\
lens & 2 (1) & 584-629 & 0.6-2.1 \\
mahout & 13 (10) & 264-906 & 0.0-10.6 \\
manifoldcf & 28 (28) & 430-1058 & 1.0-10.3 \\
nutch & 22 (22) & 245-500 & 1.7-12.5 \\
opennlp & 2 (2) & 626-632 & 0.8-1.7 \\
parquet-mr & 10 (10) & 221-429 & 2.1-9.0 \\
santuario-java & 6 (2) & 165-463 & 0.0-6.8 \\
systemml & 10 (9) & 851-1073 & 0.0-9.2 \\
tika & 28 (27) & 61-651 & 2.5-19.8 \\
wss4j & 5 (4) & 135-500 & 0.0-9.6 \\
\hline
\textbf{Total} & 398 (265) & 154.271 & 5.2
\end{tabular}
\end{table}

\subsection{Execution Plan}

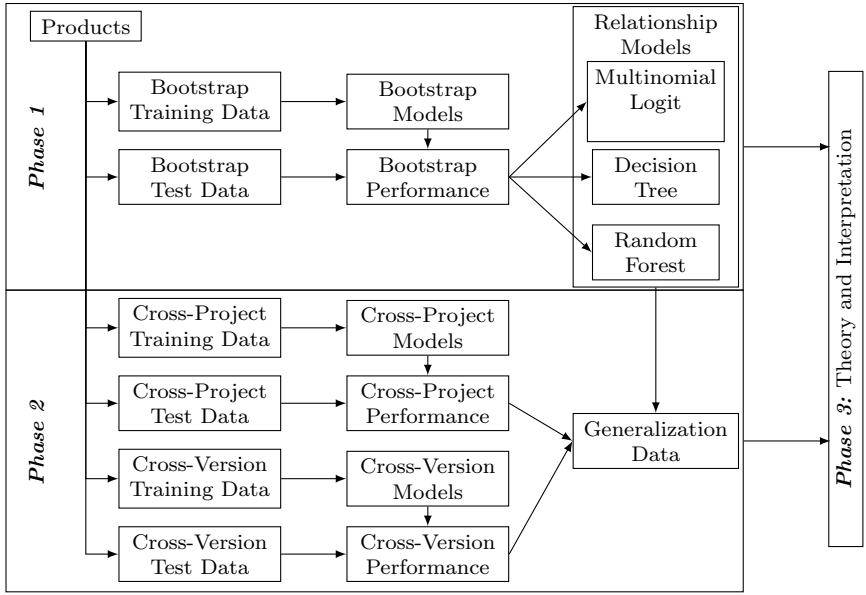
\begin{figure}
\centering
\tikzstyle{block} = [draw, text width=7.5em, text centered]
\tikzstyle{line} = [draw, >=latex']

\begin{tikzpicture}

\node[shape=rectangle,draw=black, align=center, minimum width={width("Products")+8pt}] (DATA) at (-2.5, 4) {Products};
\node[shape=rectangle,draw=black, align=center, minimum width={width("Training Data")+8pt}] (P1TR) at (-1, 3) {Bootstrap\\Training Data};
\node[shape=rectangle,draw=black, align=center, minimum width={width("Training Data")+8pt}] (P1TEST) at (-1, 2) {Bootstrap\\Test Data};
\node[shape=rectangle,draw=black, align=center, minimum width={width("Training Data")+8pt}] (P1MOD) at (2, 3) {Bootstrap\\Models};
\node[shape=rectangle,draw=black, align=center, minimum width={width("Training Data")+8pt}] (P1PF) at (2, 2) {Bootstrap\\Performance};
\node[shape=rectangle,draw=black, align=center, minimum width={width("Regression")+8pt}] (P1LR) at (5, 3) {Multinomial\\Logit\\};
\node[shape=rectangle,draw=black, align=center, minimum width={width("Regression")+8pt}] (P1DT) at (5, 2) {Decision\\Tree};
\node[shape=rectangle,draw=black, align=center, minimum width={width("Regression")+8pt}] (P1RF) at (5, 1) {Random\\Forest};

\node[draw=black, align=center, text depth = 2.99cm, minimum width={width("Generalization")+8pt},minimum height={3.7cm}] (P1RM) at (5, 2.4) {Relationship\\Models};

\draw [-latex] (P1TR.east) -- (P1MOD.west);
\draw [-latex] (P1MOD.south) -- (P1PF.north);
\draw [-latex] (P1TEST.east) -- (P1PF.west);
\draw [-latex] (P1PF.east) -- (P1LR.west);
\draw [-latex] (P1PF.east) -- (P1DT.west);
\draw [-latex] (P1PF.east) -- (P1RF.west);

\node[shape=rectangle,draw=black, align=center, minimum width={width("Training Data")+8pt}] (P2CPTR) at (-1, 0) {Cross-Project\\Training Data};
\node[shape=rectangle,draw=black, align=center, minimum width={width("Training Data")+8pt}] (P2CPTEST) at (-1, -1) {Cross-Project\\Test Data};
\node[shape=rectangle,draw=black, align=center, minimum width={width("Training Data")+8pt}] (P2CVTR) at (-1, -2) {Cross-Version\\Training Data};
\node[shape=rectangle,draw=black, align=center, minimum width={width("Training Data")+8pt}] (P2CVTEST) at (-1, -3) {Cross-Version\\Test Data};
\node[shape=rectangle,draw=black, align=center, minimum width={width("Training Data")+8pt}] (P2CPMOD) at (2, 0) {Cross-Project\\Models};
\node[shape=rectangle,draw=black, align=center, minimum width={width("Training Data")+8pt}] (P2CPPF) at (2, -1) {Cross-Project\\Performance};
\node[shape=rectangle,draw=black, align=center, minimum width={width("Training Data")+8pt}] (P2CVMOD) at (2, -2) {Cross-Version\\Models};
\node[shape=rectangle,draw=black, align=center, minimum width={width("Training Data")+8pt}] (P2CVPF) at (2, -3) {Cross-Version\\Performance};

\node[shape=rectangle,draw=black, align=center, minimum width={width("Generalization")+8pt}] (P2GD) at (5, -1.5) {Generalization\\Data};

\node[rotate=90,text depth={8.9cm},draw=black,minimum width={3.8cm},minimum height={9.7cm}] (P1) at (1.3, 2.4) {\textit{\textbf{Phase 1}}};
\node[rotate=90,text depth={8.9cm},draw=black,minimum width={4cm},minimum height={9.7cm}] (P2) at (1.3, -1.5) {\textit{\textbf{Phase 2}}};
\node[rotate=90, shape=rectangle,draw=black, minimum width=6.3cm] (P3) at (7.5, 0.25) {\textbf{\textit{Phase 3:}} Theory and Interpretation};

\draw [-latex] (DATA.south) |- (P1TR.west);
\draw [-latex] (DATA.south) |- (P1TEST.west);
\draw [-latex] (DATA.south) |- (P2CPTR.west);
\draw [-latex] (DATA.south) |- (P2CPTEST.west);
\draw [-latex] (DATA.south) |- (P2CVTR.west);
\draw [-latex] (DATA.south) |- (P2CVTEST.west);
\draw [-latex] (P2CPTR.east) -- (P2CPMOD.west);
\draw [-latex] (P2CPMOD.south) -- (P2CPPF.north);
\draw [-latex] (P2CPTEST.east) -- (P2CPPF.west);
\draw [-latex] (P2CVTR.east) -- (P2CVMOD.west);
\draw [-latex] (P2CVMOD.south) -- (P2CVPF.north);
\draw [-latex] (P2CVTEST.east) -- (P2CVPF.west);
\draw [-latex] (P1RM.south) -- (P2GD.north);
\draw [-latex] (P2CVPF.east) -- (P2GD.west);
\draw [-latex] (P2CPPF.east) -- (P2GD.west);
\draw [-latex] (P1.south) -- (7.3,2.4);
\draw [-latex] (P2.south) -- (7.3,-1.5);

\end{tikzpicture}
\caption{Overview of the execution plan. In phase 1, we create bootstrap samples from our available software products and use these samples to train the bootstrap models and collect performance data, i.e., our variables. These are then used as input to train models for the relationship between our variables. Phase 2 is similar, except that we use cross-project and cross-version defect prediction to collect performance data. We then evaluate the goodness of fit of the relationship models from phase 1 on the performance data from phase 2. The results regarding the relationships from both phases are used as input for deriving a theory and interpreting the results in phase 3.}
\label{fig:method}
\end{figure}

The execution of this study consists of three phases (Figure~\ref{fig:method}). In the first phase, we collect a large amount of data that we can use to analyze the relationship between different metrics and cost saving potential and analyze the relationship between our variables. In the second phase, we evaluate if our results generalize to other prediction settings. In the third phase, we discuss our results and try to infer a theory regarding the suitability of performance metrics as indicators for cost saving potential that can guide future work on defect prediction. 

Thus, the first phase is used to collect the data we require to build an initial model for answering our research question. We then use the data from the second phase to determine if the relationship generalizes, i.e., to determine if we found a stable relationship. The results of both phases are used for building a theory that answers our research question and that conforms to all observations and is consistent with the mathematical properties of the performance criteria and cost model.

We use CrossPare~\citep{Herbold2015} for the execution of defect prediction experiments and scikit-learn \citep{Pedregosa2011} for the modeling of our dependent variable based on the independent and confounding variables. 

\subsubsection{Bootstrapping the Relationships}

The first part of our experiments is based on results of defect prediction models obtained through bootstrap sampling. We filter the projects according to the criteria by \cite{Herbold2017b}, i.e., we use only releases with at least 100 instances and five defects. This ensures that our bootstrap experiment does not include very small releases or releases with too few bugs to build reasonable predictors. This leaves us with 265 releases. We draw 100 bootstrap samples from each release. We draw again in case the bootstrap sample does not contain at least two defects in the training data and one defect in the test data. This is a small deviation from the registered protocol, where we only required one defect. The reason for this change is that SMOTUNED~\citep{Agrawal2018} requires at least two defective instances for training. We train on the instances in the bootstrap samples and test on the remaining instances. According to \cite{Tantithamthavorn2017}, this should lead to an unbiased estimator of the performance means. Based on the work by \cite{Hosseini2017} and \cite{Tantithamthavorn2016}, we decided to use a random forest~\citep{Breiman2001} with hyperparameter optimization as classifier. Table~\ref{tbl:rfparams} lists the ranges from which the hyperparameters are sampled using differential evolution~\citep{Qing2009}. We train two variants of each classifier. The first gets the training data as is, i.e., without any modifications. For the second, we use SMOTUNED oversampling to mitigate the class level imbalance~\citep{Agrawal2018}. We measure all variables by applying the trained classifier to the test data. 

\begin{table}
\centering
\caption{Configuration of the hyperparameter selection. The tree depth is not limited. The ranges are based on a library for automated parameter optimization~\cite{Feurer2015}.}
\label{tbl:rfparams}
\begin{tabular}{llll}
\textbf{Parameter} & \textbf{type} &  \textbf{min} & \textbf{max} \\
\hline\hline
Ratio of features available for each tree & float & 0 & 1 \\
Minimal instances for a decision & integer & 2 & 20 \\
Minimal instances for a leaf & integer & 1 & 20 \\
\hline
\end{tabular}
\end{table}

We have 100 bootstrap results without treatment of class level imbalance and 100 bootstrap results with treatment of the class level imbalance for each release in the data. Thus, we have $(100+100)\cdot 265 = 53,000$ instances which we can use for the analysis of the relationship between metrics. We explore the observed values of the cost boundaries \textit{diff}. We visualize the observed values through a histogram. This plot does not contain any values for \textit{diff} that are \textit{NaN} or $\pm \infty$. We report how often these corner cases occur separately. This data is the core of our sensitivity analysis (Section~\ref{sec:sensitivity}). We report Spearman's rank correlation~\citep{spearman1961proof} between all pairs of variables. The reporting of correlations is an extension of our registered protocol that allows us to incorporate the interactions between the independent and confounding variables in our analysis.

We use three models to analyze the relationship between the independent variables and the dependent variable: 1) a multinomial logit model to get insights into the changes in likelihood for cost savings, given the performance metrics; 2) a decision tree to get a different view on the relationship that is based on decision rules; and 3) a random forest as powerful non-linear model that we use to determine if the commonly used metrics are suitable proxies for the cost efficiency of defect prediction models. 

The multinomial logit model is used to model the linear relationship of the log-odds of the levels of the dependent variable and the independent and confounding variables. We first train a model with elastic net regression, i.e., we use a combination of ridge regularization to avoid large coefficients that may be the result of correlations and lasso regularization to select relevant metrics. We sample the regularization strength exponentially from $10^{0}, 10^{1}, ..., 10^5$ and the ratio between ridge and lasso linearly from $0.0, 0.1, ..., 1.0$. We maximize the adjusted $R^2$ statistic to determine suitable a regularization strength. Adjusted $R^2$ increases if the logit model explains a larger proportion of the variance of the data and decreases with more model parameters. Hence, we find a good trade-off between the model complexity and the goodness of fit if we select the regularization parameters to optimize the adjusted $R^2$ statistics. Since we have an ordinal dependent variable, we use McFadden's adjusted $R^2$~\citep{mcfadden1974conditional}, which is defined as
\begin{equation}
    R^{2} = 1-\frac{(\sum_{i=1}^m y_i^* \cdot \ln y_i)-k}{\sum_{i=1}^m y_i^* \cdot \ln y_i^{NULL}}
\end{equation}
where $k$ is the number of independent and confounding variables used, $y_i$ are the dependent variables, $y_i^*$ is the expected outcome, i.e., the actual value of the \textit{potential} in our case, $y_i$ is the predicted score by the multinomial logit model, and $y_i^{NULL}$ is the score of a multinomial logit without variables, i.e., with only an intercept.\footnote{We use a one-hot encoding for the dependent variables here.} For this model, we normalize all independent and confounding variables because the regularization would otherwise not work correctly. We use this first multinomial logit model for the selection of uncorrelated features. Afterwards, we train a second model without normalization, using only the features that were previously selected. Since the normalization does not affect the model quality, but only the regularization, we train this second model without regularization, assuming that we already have an uncorrelated set of features. 

First training with normalization for finding an uncorrelated subset with the elastic net, followed by training without regularization and normalization is a deviation from our pre-registered protocol. However, this was required because some of our variables have very different scales. While many performance metrics are within the interval $[0, 1]$, other variables, like $N_{train}$ are on much larger scales. This would lead to big differences in the absolute size of coefficients, i.e., variables on the smaller scale would have larger coefficients. Since the regularization penalizes coefficient sizes, this would bias the selection of variables. However, we cannot, in general, normalize the data because we would have to account for the normalization when interpreting the coefficients of the model. 

Note that we do not include p-values for the coefficients of the multinomial logit model because we are using a regularized regression to address collinearity. P-values and confidence intervals for regularized linear models are usually not reported as the introduced bias via regularization is not trivial to estimate \citep{lockhart2014significance}.

The decision tree model provides us with an easy to interpret visual model for the analysis of the relationship between performance metrics and cost saving potential. We use a CART decision tree~\citep{Breiman1984} with the Gini impurity as splitting criterion. We limit the depth of the decision tree to five, which means that the resulting decision tree has at most $2^5=32$ leaf nodes. This restriction limits the complexity of the resulting tree and ensures that manual analysis is feasible. We analyze the decision rules within the tree structure to determine how the different variables interact when forming a decision about the cost saving potential. Moreover, the decision tree allows us to determine the importance of each metric using the feature importance for the decision tree model. The feature importance measures how much each feature contributed to the reduction of the Gini impurity that is observed at the leaf nodes of the tree. Thus, a high feature importance indicates that a performance metric is well suited to determine the cost saving potential as it can reduce the uncertainty regarding the outcome. 

The random forest is consistently among the best performing machine learning algorithms~\citep{FernandezDelgado2014}.\footnote{Deep learning is not an option because we neither have enough features, nor enough data.} Random forests provide a powerful non-linear model for classification problems by averaging over many decision trees that were trained on subsets of the data and features. Same as above, we optimize the hyperparameters of the random forest based on the ranges reported in Table~\ref{tbl:rfparams}. While random forests are in principle a black box and it is unclear how decisions are made, we can still gain insights regarding the feature importance the same way as for the decision tree model. For this, the average feature importance over all trees is considered. 

We consider the confusion matrix for all three models to evaluate the goodness of fit. In our case, the confusion matrix is a $4 \times 4$ matrix, where the columns contain the true label (\textit{none}, \textit{medium}, \textit{large}, \textit{extra large}) and the rows the predicted label. Thus, the confusion matrices give us detailed insights into the models. The advantage of using the confusion matrix directly instead of metrics based on the confusion matrix is that no information is lost or hidden due to aggregation. This is especially important because we do not have a binary problem with two classes, but rather a problem with four classes. The distinction if the models are a good fit only for few classes or for all classes, is very important to understand our data and to correctly interpret how well-suited the independent variables are as proxy for costs. In particular, we evaluate the confusion matrix with respect to the following criteria.

\begin{itemize}
\item The correctly predicted instances for each class. 
\item The instances that are predicted in the upper neighbor of a class to check for a tendency of moderate overprediction. We also check how many instances are predicted in any of the classes with more cost saving potential to check for the overall overprediction. 
\item The instances that are predicted in the lower neighbor of a class to check for a tendency of moderate underprediction. We also check how many instances are predicted in any of the classes with less cost saving potential, to check for the overall underpredictions of that class. 
\end{itemize}

This gives us detailed insights into how accurate the predictions are for each level, as well as the mistakes that the model makes. 

\subsubsection{Generalization to Realistic Settings}

We apply the models trained on the bootstrap results to new data to test the generalizability of the findings. We train six defect prediction models to generate data. Three models for cross-version defect prediction and three models for strict cross-project defect prediction. We use two recent benchmarks for the selection of these models \citep{Amasaki2020, Herbold2017d}. We select the best, median, and worst ranking model. This should allow us to see how well the models about the relationship between our variables generalize to realistic defect prediction models of different quality. We get the following six models. 

\begin{itemize}
\item Cross-version defect prediction:
\begin{itemize}
    \item The training data as is, without any modifications, i.e., one of the baselines used by \cite{Amasaki2020}. 
    \item \cite{Kawata2015} propose to use DBSCAN to select a suitable subset of training data. Random forest is used as classifier. 
    \item \cite{Peters2015} proposed LACE2 as an extension of the CLIFF instance selection and the MORPH data privatization approaches~\citep{Peters2013}. Naive Bayes is used as classifier. 
\end{itemize}
\item Cross-project defect prediction:
\begin{itemize}
    \item The MODEP multi-objective genetic program suggested by \cite{Canfora2013}. Same as \cite{Herbold2017b}, we select the best classifier such that the \textit{recall} is at least 0.7. 
    \item \cite{Watanabe2008} suggest to standardize the training data based on the mean value of the target project. Naive Bayes is used a classifier. 
    \item An approach suggested by \cite{CamargoCruz2009} that proposes to apply the logarithm to all features and then standardize the training data based on the median of the target project. Naive Bayes is used as classifier. 
\end{itemize}
\end{itemize}

Same as \cite{Amasaki2020}, we use the data from the prior release for training because \cite{Amasaki2020} found that this was the best scenario for cross-version defect prediction. Same as \cite{Herbold2017b}, we use all data that is not from the target project as training data. In both cases, we discard all releases that do not have at least 100 instances or five defects from the training data. For cross-version defect prediction, this means that we do not always use the prior release, but rather the first prior release that meets these criteria. We do not apply this criterion for the filtering of the test data, i.e., we allow all projects as test data. Our rationale for this decision is that it is possible to filter extreme data for training. However, in practice it is impossible to avoid extreme data or rule out that the defect prediction model is applied to small projects or in settings with only few bugs. 

In comparison to the prior benchmark studies, we clean the available training data from any information leakage due to temporal dependencies~\citep{Bangash2020}. The MYNBOU data contains the timestamps of the releases as well as for the date of the bug fixes. We use this information to remove all data that was not available at the time of the release of the project for which the model is trained from the available training data. Additionally, we remove all releases that are closer than 6 months to the release of the target project from the training data used for the cross-project experiments. Our rationale for this additional filtering is that there would not have been sufficient time to report and fix bugs for those releases otherwise, which would mean that there could only be very few -- if any -- bugs within those data sets. Thus, the data from these releases would be unreliable and should not be used. We do not apply this criterion to the cross-version experiments because we assume that the gap between releases is sufficiently large to take care of this problem. 

We measure the dependent, independent, and confounding variables in the same way as for the bootstrap experiment. We then evaluate the confusion matrices that we get when we apply the models trained on the bootstrap results to this new data. As an extension of our registered protocol, we also report the distributions and correlations among the independent and confounding variables. 

\subsubsection{Interpretation and Theory Building}
\label{sec:theory}

Within the final phase of our study, we take all results into account to formulate a theory regarding the relationship between performance metrics and cost saving potential. There are two principle outcomes that could shape the structure of the theory. 

\begin{itemize}
\item We cannot establish a strong relationship between our variables. We observe this through the confusion matrices in both the bootstrap experiment and the generalization to other defect prediction data. We analyze the results in detail to determine why we could not establish such a theory. In case we find a significant flaw in our methodology through this analysis, we outline how future experiments could avoid this issue and, thereby, at least contribute to the body of knowledge regarding case study guidelines. However, due to the rigorous review of our experiment protocol through the registration, we believe that it is unlikely that we find such a flaw. In case we find no flaw in the methodology, we try to determine the reasons why the metrics are not suitable proxies and try to infer if similar problems may affect other machine learning applications in software engineering. We look for reasons for this lack of a relationship both through analytic considerations of the relationship between the performance metrics and the costs, as well as due to possible explanations directly within the data.
\item We can establish a strong relationship between our variables. We observe this at least through the confusion matrices in the bootstrap experiment, but possibly not when we evaluate the generalization to other models. We use the insights from the multinomial logit, decision tree, and random forest regarding the importance of the independent and confounding variables and how they contribute to the result. We combine these insights to understand which combination of variables is suited for the prediction of the cost saving potential and can, therefore, be used as suitable proxy. We derive a theory regarding suitable proxies from these insights and how they should be used. This theory includes how the proxies are mathematically related to the cost to understand the causal relationships that lead to the criteria being good proxies. The theory may also indicate that there are no suitable proxies, in case the confounding variables are key drivers of the prediction of cost saving potential. We would interpret this as strong indication that cost saving potential depends on the structure of the training and/or test data and cannot be extrapolated from performance metrics.
\end{itemize}

We may identify different types of ``strong relationships''. We specify the following potential results. 

\begin{itemize}
\item Cost saving potential classification possible: 
\begin{itemize}
\item At least 90\% of instances that are not cost saving are predicted correctly (level \textit{none}).
\item At least 90\% of instances that have cost saving potential are predicted correctly (not in level \textit{none}). 
\end{itemize}
\item Cost saving potential categorization possible (weak):
\begin{itemize}
\item The two criteria above are fulfilled.
\item At least 90\% of the instances with cost saving potential are either in the correct level or in a neighboring cost saving level. For example, 90\% of instances with \textit{medium} cost saving potential are either predicted as \textit{medium} or \textit{or large}.
\end{itemize}
\item Cost saving potential categorization possible (strong):
\begin{itemize}
\item At least 90\% of instances of each level are predicted correctly. 
\end{itemize}
\end{itemize}

Thus, our theory accounts for the strength of the results, both with respect to the capability to distinguish between not cost saving at all or possibly cost saving as well as the ability to identify the cost saving potential. If none of the conditions are fulfilled, we conclude that we did not find a strong relationship. 

\subsubsection{Sensitivity Analysis}
\label{sec:sensitivity}

The levels for our ordinal independent variable \textit{potential} may affect the results. The biggest risk is that many values of \textit{diff} are close to the boundary. For example, the absolute difference between $\textit{diff}=1000$ and $\textit{diff}=1001$ is negligible, but the first one has a \textit{medium potential} and the second one a \textit{large potential}. If this is often the case, the choice of an ordinal independent variable may have a negative effect on our results such that we underestimate the capability for strong categorization, which requires us to predict the levels correctly from the performance metrics. The weak categorization does not have the same risk because the requirement is relaxed such that prediction is in a neighboring level are sufficient. The classification is also not affected because this ignores the choice of levels altogether and reduces the cost saving potential to a binary problem.

We explore this risk by exploring the distribution of \textit{diff} with a focus on the values that are within 10\% of the boundary values. For example, the boundary value between \textit{medium} and \textit{large} is 1000. Hence, we explore the interval $[900,1100]$. For the boundary between \textit{large} and \textit{extra large} the boundary value is 10000, hence, we explore the interval $[9000, 11000]$. This way, we take the exponential growth of the ranges of the levels into account. If we observe that the distribution indicates that an over proportionally large amount of data is within these relatively small regions, i.e. more than 20\% of the data for a level, this means that our boundaries are within dense regions which would make the strong categorization harder. In this case, we further explore if strong categorization is possible with a random forest, if we shift the boundaries by 10\% upwards and downwards. This means we build two more random forest models, one where the boundaries are at $0.9 \cdot 10^n$ (downwards shift) and one where the boundaries are at $1.1 \cdot 10^n$ (upwards shift) with $n=1, ..., 5$. We take the results of this sensitivity analysis into account for the interpretation and generation of our theory.

A second and smaller risk is that our choice to use an ordinal dependent variable may hide an existing relationship between the dependent and confounding variables and the independent variable which we could observe through a regression model of $\textit{diff}$. This should only be the case if we do not observe strong or weak categorization due to completely inappropriate levels, e.g. because categorization would be possible with linearly distributed boundaries instead of the exponential growth we model. However, this risk is negligible because the random forest we use is invariant towards monotonic transformations of the independent variables, i.e., only the order of values matters. While this statement technically applies only to the random forest regression, this also holds true without limitations for the prediction of a ordinal variable with a random forest classifier that is defined based on boundaries for the variable that would be the target of the regression model. Thus, if the random forest fails to model the exponentially distributed boundaries, there is no reason to believe that this works better for boundaries that follow a different distribution, since this means that the random forest is not able to correctly approximate the order of the instances. Consequently, we do not conduct an additional sensitivity analysis regarding this choice, other than to confirm that this is indeed the case.

\subsection{Summary of Deviations}
\label{sec:deviations}

In summary, we have the following deviations from our registered research protocol. 

\begin{itemize}
    \item We modified our dependent variable to only have four levels instead of six, due to data scarcity in the other two levels. 
    \item We strengthened the requirements on the data for the bootstrap experiment to enforce that there are at least two defective artifacts to ensure that SMOTUNED~\citep{Agrawal2018} is always usable. 
    \item We calculate Spearman's rank correlation~\citep{spearman1961proof} between all independent and confounding variables to enable the consideration of the interactions between variables in our theory building.
    \item We first train a multinomial logit model with normalized variables to select the relevant variables through regularization and subsequently train the model on the selected variables without regularization for the analysis.
\end{itemize}

Moreover, we clarified some aspects which were yet fully defined in the research protocol.

\begin{itemize}
    \item We use CrossPare~\citep{Herbold2015} for the execution of all defect prediction experiments and scitkit-learn~\citep{Pedregosa2011} for the analysis of the dependent variable based on the independent and confounding variables. 
    \item We specified that we used differential evolution~\cite{Qing2009} for the hyperparameter tuning of the random forests and a grid search for the multinomial logit model.
    \item We specified that we use McFadden's adjusted $R^2$~\citep{mcfadden1974conditional} for the selection of the best hyperparameters of the multinomial logit model because we have an ordinal dependent variable.
\end{itemize}

\section{Results}
\label{sec:results}

We now present the results of our experiments. First, we present the bootstrap experiment, then the generalization experiment, where we use cross-version and cross-project defect prediction. Finally, we discuss the sensitivity of our results to the choice of bins for our dependent variable. We share our results and code produced as part of this experiment through our replication package.\footnote{https://github.com/sherbold/replication-kit-defect-prediction-metrics}

\subsection{Bootstrap Experiment}
\label{sec:bootstrap}

\begin{figure}
\centering
\includegraphics[width=\textwidth]{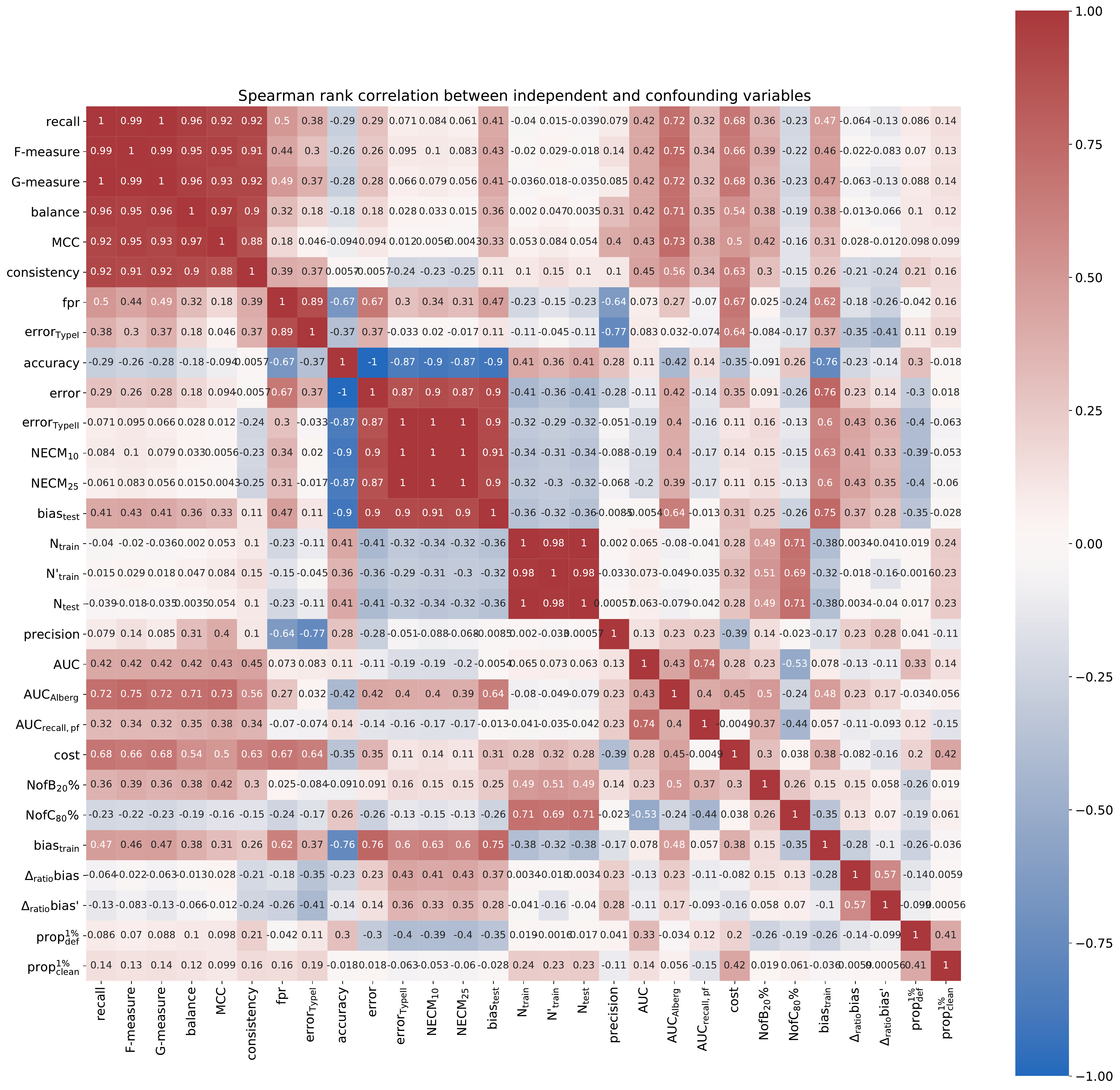}
\caption{Correlations between the variables (independent and confounding) in the bootstrap experiment. The variables are ordered to show strong correlations close to the diagonal to simplify the identification of groups of correlated features.}
\label{fig:correlations}
\end{figure}

We found four groups of strongly correlated variables ($>$0.8), which are depicted in Figure~\ref{fig:correlations}. 
\begin{itemize}
    \item the \textit{recall} group with the metrics \textit{recall, F-measure, G-measure, balance, MCC}, and \textit{consistency};
    \item the \textit{fpr} group with the metrics \textit{fpr} and $\textit{error}_\textit{TypeI}$;
    \item the \textit{accuracy} group with the metrics \textit{accuracy}, \textit{error}, $\textit{error}_\textit{TypeII}$, $\textit{NECM}_{10}$, $\textit{NECM}_{25}$, and $\textit{bias}_{\textit{test}}$; and
    \item the $N$ group with the metrics $N_\textit{train}$, $N'_\textit{train}$, and $N_\textit{test}$.
\end{itemize}

The strong correlations are to some degree expected, but also contain some surprises. For example, the \textit{F-measure} as the harmonic mean of \textit{recall} and \textit{precision} is only correlated to \textit{recall}, but not \textit{precision}. We note that there are additional correlations of medium strengths, e.g., between \textit{AUC} and $\textit{AUC}_\textit{recall,pf}$, between $\textit{AUC}_\textit{Alberg}$ and the \textit{recall} group, and between $\textit{bias}_\textit{train}$ and the \textit{accuracy} group.

\begin{figure}
\centering
\includegraphics[width=\textwidth]{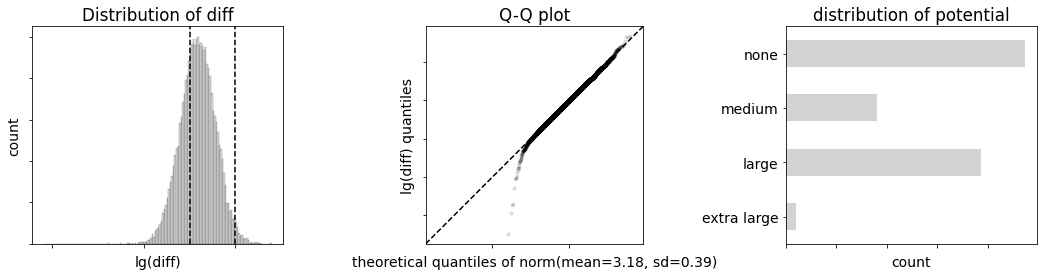}
\caption{Distribution of \textit{diff} and \textit{potential} in the bootstrap experiment. The left-most plot shows a histogram of the decadic logarithm ($lg$) of \textit{diff}, the middle shows a Q-Q-plot of $lg(\textit{diff})$ and a normal distribution with $mean=3.18$ and $sd=0.39$. The plots of diff ignore 30 negative values, 583 values of $+\infty$, 11,521 values of $-\infty$ and 12,075 \textit{NaN}s. The right-most plot shows the distribution of the dependent variable \textit{potential}.}
\label{fig:dependent_bootstrap}
\end{figure}

Figure~\ref{fig:dependent_bootstrap} shows the distributions of our dependent variable \textit{potential} and of the variable \textit{diff}, on which the \textit{potential} is based. The data shows that \textit{diff} is almost perfectly normally distributed, when considered transformed with the decadic logarithm. Moreover, the data shows that we have many values which would be problematic if we were to work with \textit{diff} directly, i.e., $\pm\infty$ and \textit{NaN}. Overall, these distributions support our choice of \textit{potential} as ordinal variable with powers of ten as bins as our dependent variable. The large number of \textit{NaN} values is due to models that do not predict any defects on the test data. The \textit{NaN} occur both without SMOTUNED (n=7890) and with SMOTUNED (n=4185). Moreover, while the number of defects in the test data is not large, it is also not about finding a single ``bug in a haystack'', with a median of 5 bugs that could be found. In our opinion, better than random and non-trivial predictors should be able to find at least one of these bugs. Thus, while this is a large number of \textit{NaN}s, this is not an artifact of our experiment design, but more likely due to the actually bad quality of the prediction models. 

\begin{figure}
\centering
\includegraphics[width=\textwidth]{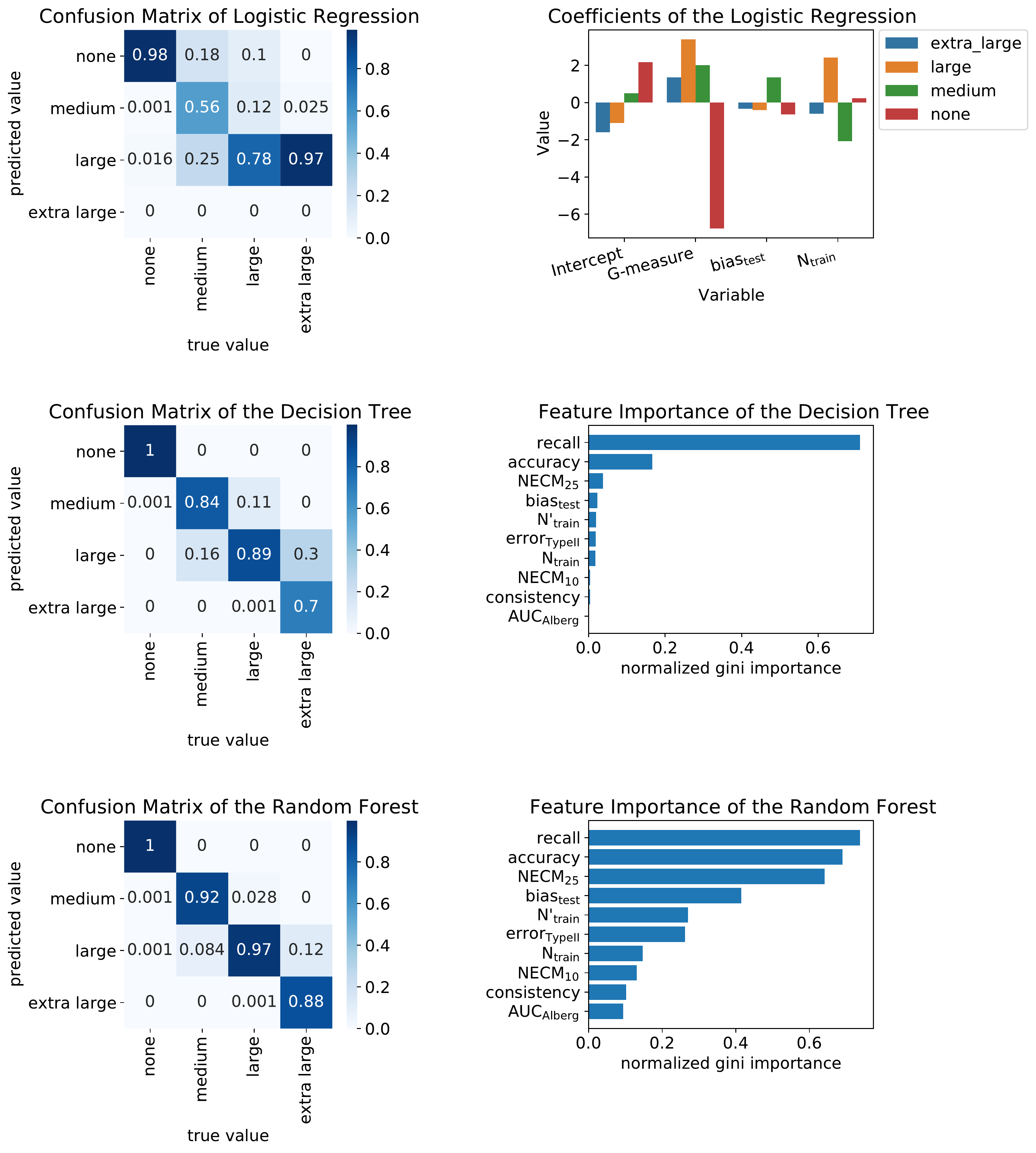}
\caption{Confusion matrices and features of the bootstrap experiment. For the multinomial logit model, we show the non-zero coefficients of the model. The coefficients for $N_\textit{train}$ are multiplied with 1000, which is roughly the difference between the value ranges of the variables for which we have coefficients.}
\label{fig:phase1_confmat}
\end{figure}

Figure~\ref{fig:phase1_confmat} shows an overview of the results of the first phase of the experiments, i.e., the training of predictors of the \textit{potential} based on the bootstrap experiment. The grid search determined that a Lasso model with a regularization strength of 1000 is optimal for the multinomial logit model. The resulting model only uses three variables: \textit{G-measure}, $\textit{bias}_\textit{test}$ and $N_\textit{train}$. The resulting model only predicts three classes, i.e., \textit{none}, \textit{medium}, and \textit{large}. The confusion matrix shows that the data from the class \textit{extra large} is nearly always predicted as \textit{large}. The class \textit{None} is almost always correctly identified. The class \textit{medium} is predicted correctly in about half of the cases, the wrong predictions of \textit{medium} do not follow a clear pattern and fall into both classes \textit{none} and \textit{large}. For \textit{large}, about three of four predictions are correct and there is also no clear pattern for these mistakes, which fall into both \textit{none} and \textit{medium}.

\begin{figure}
\centering
\tikzstyle{block} = [draw, text width=7.5em, text centered]
\tikzstyle{line} = [draw, >=latex']

\begin{tikzpicture}
\node[shape=rectangle,draw=black, align=left] (0) at (0, 0) {$\textit{recall} \leq 0.01$\\1041, 19332, 9001, 23626\\class = none};
\node[shape=rectangle,draw=black, align=left, minimum width={width("Products")+8pt}] (1) at (3, 0) {0, 0, 0, 23596\\class = none};
\node[shape=rectangle,draw=black, align=left, minimum width={width("Products")+8pt}] (2) at (0, -2) {$\textit{accuracy} \leq 0.898$\\1041, 19332, 9001, 30\\class = large};
\node[shape=rectangle,draw=black, align=left, minimum width={width("Products")+8pt}] (3) at (3, -2) {$N_\textit{train} \leq 527.5$\\75, 2003, 6510, 17\\class = medium};
\node[shape=rectangle,draw=black, align=left, minimum width={width("Products")+8pt}] (4) at (6, -2) {$\textit{error}_\textit{TypeII} \leq 0.049$\\75, 1206, 6100, 14\\class = medium};
\node[shape=rectangle,draw=black, align=left] (5) at (9, -2) {$\textit{consistency} \leq 0.944$\\75, 492, 169, 0\\class = large};
\node[shape=rectangle,draw=black, align=left, minimum width={width("Products")+8pt}] (6) at (12, -1.5) {1, 492, 169, 0\\class = large};
\node[shape=rectangle,draw=black, align=left, minimum width={width("Products")+8pt}] (7) at (12, -2.5) {74, 0, 0, 0\\class = extra large};
\node[shape=rectangle,draw=black, align=left, minimum width={width("Products")+8pt}] (8) at (9, -4) {$\textit{bias}_\textit{test} \leq 0.099$\\0, 714, 5931, 14\\class = medium};
\node[shape=rectangle,draw=black, align=left, minimum width={width("Products")+8pt}] (9) at (12, -3.5) {0, 370, 747, 0\\class = medium};
\node[shape=rectangle,draw=black, align=left, minimum width={width("Products")+8pt}] (10) at (12, -4.5) {0, 344, 5184, 14\\class = medium};
\node[shape=rectangle,draw=black, align=left, minimum width={width("Products")+8pt}] (11) at (6, -6)
{$\textit{bias}_\textit{test} \leq 0.108$\\0, 797, 410, 3\\class = large};
\node[shape=rectangle,draw=black, align=left, minimum width={width("Products")+8pt}] (12) at (9, -6)
{$\textit{accuracy} \leq 0.881$\\0, 421, 79, 0\\class = large};
\node[shape=rectangle,draw=black, align=left, minimum width={width("Products")+8pt}] (13) at (12, -5.5) {0, 48, 26, 0\\class = large};
\node[shape=rectangle,draw=black, align=left, minimum width={width("Products")+8pt}] (14) at (12, -6.5) {0, 373, 53, 0\\class = large};
\node[shape=rectangle,draw=black, align=left, minimum width={width("Products")+8pt}] (15) at (9, -8)
{$\textit{NECM}_{10} \leq 1.084$\\0, 376, 331, 3\\class = large};
\node[shape=rectangle,draw=black, align=left, minimum width={width("Products")+8pt}] (16) at (12, -7.5) {0, 324, 241, 3\\class = large};
\node[shape=rectangle,draw=black, align=left, minimum width={width("Products")+8pt}] (17) at (12, -8.5) {0, 52, 90, 0\\class = medium};
\node[shape=rectangle,draw=black, align=left, minimum width={width("Products")+8pt}] (18) at (3, -10)
{$\textit{NECM}_{25} \leq 0.178$\\966, 17329, 2491, 13\\class = large};
\node[shape=rectangle,draw=black, align=left, minimum width={width("Products")+8pt}] (19) at (6, -10)
{$\textit{error}_\textit{TypeII} \leq 0.007$\\659, 27, 0, 0\\class = extra large};
\node[shape=rectangle,draw=black, align=left, minimum width={width("Products")+8pt}] (20) at (9, -10)
{$N_\textit{train} \leq 1033.5$\\626, 8, 0, 0\\class = extra large};
\node[shape=rectangle,draw=black, align=left, minimum width={width("Products")+8pt}] (21) at (12, -9.5) {626, 7, 0, 0\\class = extra large};
\node[shape=rectangle,draw=black, align=left, minimum width={width("Products")+8pt}] (22) at (12, -10.5) {0, 1, 0, 0\\class = large};
\node[shape=rectangle,draw=black, align=left, minimum width={width("Products")+8pt}] (23) at (9, -12)
{$\textit{AUC}_\textit{Alberg} \leq 0.025$\\33, 19, 0, 0\\class = extra large};
\node[shape=rectangle,draw=black, align=left, minimum width={width("Products")+8pt}] (24) at (12, -11.5) {32, 9, 0, 0\\class = extra large};
\node[shape=rectangle,draw=black, align=left, minimum width={width("Products")+8pt}] (25) at (12, -12.5) {1, 10, 0, 0\\class = large};
\node[shape=rectangle,draw=black, align=left, minimum width={width("Products")+8pt}] (26) at (6, -14)
{$\textit{bias}_\textit{test} \leq 0.078$\\307, 17302, 2491, 13\\class = large};
\node[shape=rectangle,draw=black, align=left, minimum width={width("Products")+8pt}] (27) at (9, -14)
{$\textit{NECM}_{10} \leq 0.131$\\307, 13441, 799, 5\\class = large};
\node[shape=rectangle,draw=black, align=left, minimum width={width("Products")+8pt}] (28) at (12, -13.5) {193, 355, 0, 0\\class = large};
\node[shape=rectangle,draw=black, align=left, minimum width={width("Products")+8pt}] (29) at (12, -14.5) {114, 13086, 799, 5\\class = large};
\node[shape=rectangle,draw=black, align=left, minimum width={width("Products")+8pt}] (30) at (9, -16)
{$N'_\textit{train} \leq 628.5$\\0, 3861, 1692, 8\\class = large};
\node[shape=rectangle,draw=black, align=left, minimum width={width("Products")+8pt}] (31) at (12, -15.5) {0, 1429, 1517, 8\\class = medium};
\node[shape=rectangle,draw=black, align=left, minimum width={width("Products")+8pt}] (32) at (12, -16.5) {0, 2432, 175, 0\\class = large};

\draw [-latex] (0.east) -- (1.west);
\draw [-latex] (0.south) -- (2.north);
\draw [-latex] (2.east) -- (3.west);
\draw [-latex] (3.east) -- (4.west);
\draw [-latex] (4.east) -- (5.west);
\draw [-latex] (5.east) -- (6.west);
\draw [-latex] (5.east) -- (7.west);
\draw [-latex] (4.south) |- (8.west);
\draw [-latex] (8.east) -- (9.west);
\draw [-latex] (8.east) -- (10.west);
\draw [-latex] (3.south) |- (11.west);
\draw [-latex] (11.east) -- (12.west);
\draw [-latex] (12.east) -- (13.west);
\draw [-latex] (12.east) -- (14.west);
\draw [-latex] (11.south) |- (15.west);
\draw [-latex] (15.east) -- (16.west);
\draw [-latex] (15.east) -- (17.west);
\draw [-latex] (2.south) |- (18.west);
\draw [-latex] (18.east) -- (19.west);
\draw [-latex] (19.east) -- (20.west);
\draw [-latex] (20.east) -- (21.west);
\draw [-latex] (20.east) -- (22.west);
\draw [-latex] (19.south) |- (23.west);
\draw [-latex] (23.east) -- (24.west);
\draw [-latex] (23.east) -- (25.west);
\draw [-latex] (18.south) |- (26.west);
\draw [-latex] (26.east) -- (27.west);
\draw [-latex] (27.east) -- (28.west);
\draw [-latex] (27.east) -- (29.west);
\draw [-latex] (26.south) |- (30.west);
\draw [-latex] (30.east) -- (31.west);
\draw [-latex] (30.east) -- (32.west);

\end{tikzpicture}
\caption{Decision Tree trained generated as part of the bootstrap experiment. Each node shows the decision that is made (first row), the distribution of the classes (second row, from \textit{extra large} to \textit{none}), and the most likely class for the node (third row).}
\label{fig:phase1_dt}
\end{figure}
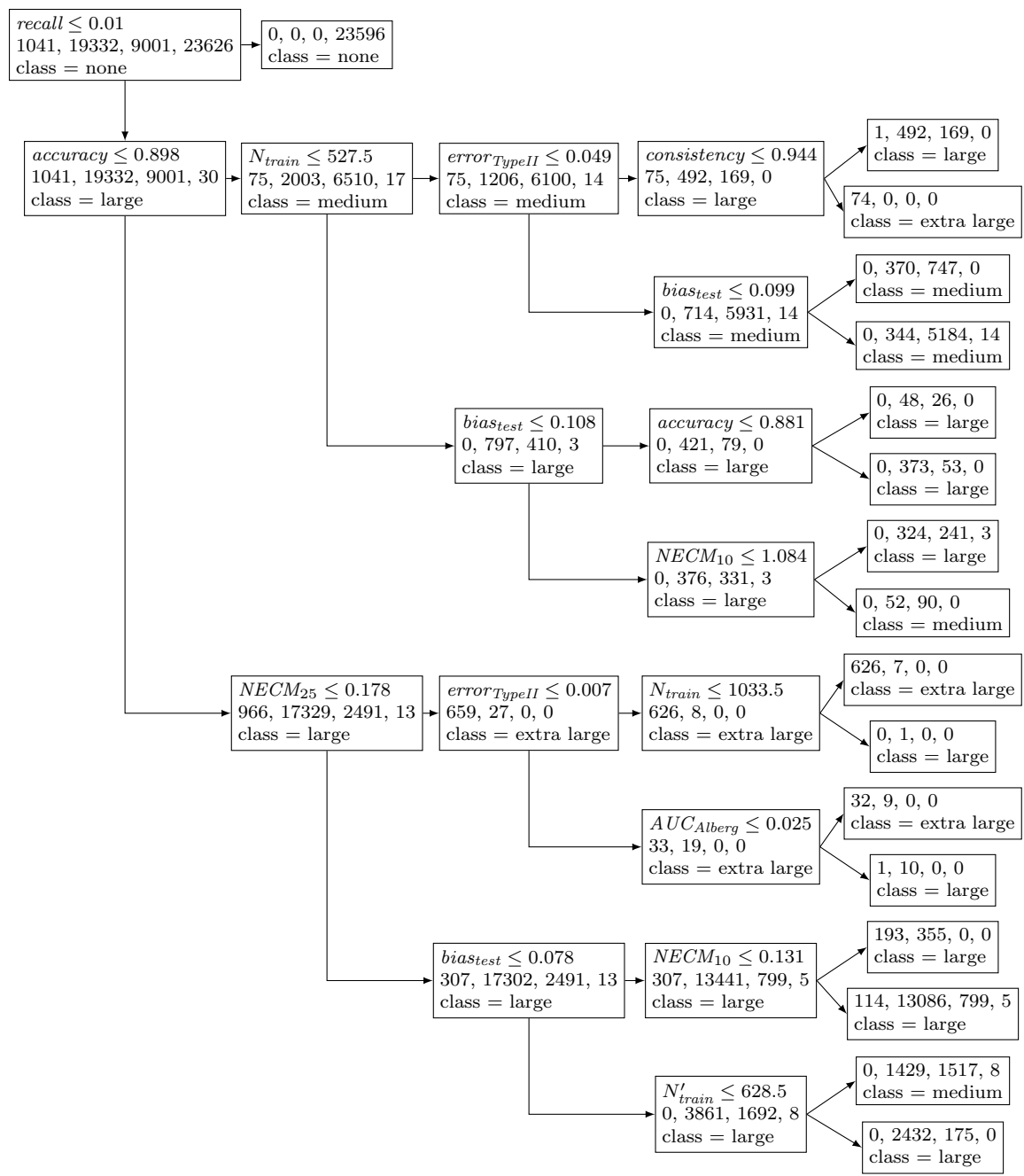

The decision tree also has an almost perfect separation between the \textit{none} and the other classes. However, in comparison to the multinomial logit model, the confusion matrix shows  that the three classes \textit{medium}, \textit{large}, and \textit{extra large} are all predicted fairly accurately, i.e., with at least 70\% of instances correct for each class. All wrong predictions of the cost saving classes are in their neighbors, e.g., while 16\% of the \textit{medium} instances are predicted as \textit{large}, they are never predicted as \textit{extra large}. Consequently, all errors are moderate over and underprediction. When we consider the feature importance (Figure~\ref{fig:phase1_confmat}) and the structure of the decision tree (Figure~\ref{fig:phase1_dt}), we observe that the \textit{recall} plays an important role, i.e., if the \textit{recall} is less than or equal to 0.01, the \textit{potential} is \textit{none}. This explains all but 30 instances, with the class \textit{none}, including all \textit{NaN}s. The remaining 30 instances are not classified correctly. This mechanism is similar to how the multinomial logit model because \textit{G-measure} and \textit{recall} are strongly correlated. The \textit{accuracy}, as the second most important variable, is the main driver to distinguish \textit{medium} from \textit{large} and \textit{extra large}. We note that \textit{accuracy} has a strong negative correlation with the confounding variable $\textit{bias}_\textit{test}$. We also observe variables from the \textit{accuracy} group are used for seven more decisions about the \textit{potential}, including $\textit{bias}_\textit{test}$, which is used thrice. The variable $\textit{NECM}_{25}$ as main driver to distinguish \textit{large} from \textit{extra large} is also from the \textit{accuracy} group. We also note that the confounding variables from the $N$ correlation group are used by the tree to partition the data such that larger projects have a lower cost saving potential.\footnote{One decision does not support this ($N_\textit{train}\leq 1033.5$). However, this decision is clearly overfitting, as this is used to separate a single project into a leaf node.}

The results of the random forest are mostly in line with what we observed with the decision tree, except that the number of errors for all classes is reduced, such that at least 88\% are correct for each class. The feature importance aligns with the results from the decision tree, if we account for the correlation group: features from the \textit{recall} group are at the top, followed by features from the \textit{accuarcy} group. This indicates that similar decision structures are used, i.e., the \textit{recall} group for the differentiation between cost saving and not cost saving and the \textit{accuracy} group for the differentiation between \textit{medium}, \textit{large}, and \textit{extra large}. 

\subsection{Generalization}
\label{sec:generalization}

\begin{figure}
\centering
\includegraphics[width=\textwidth]{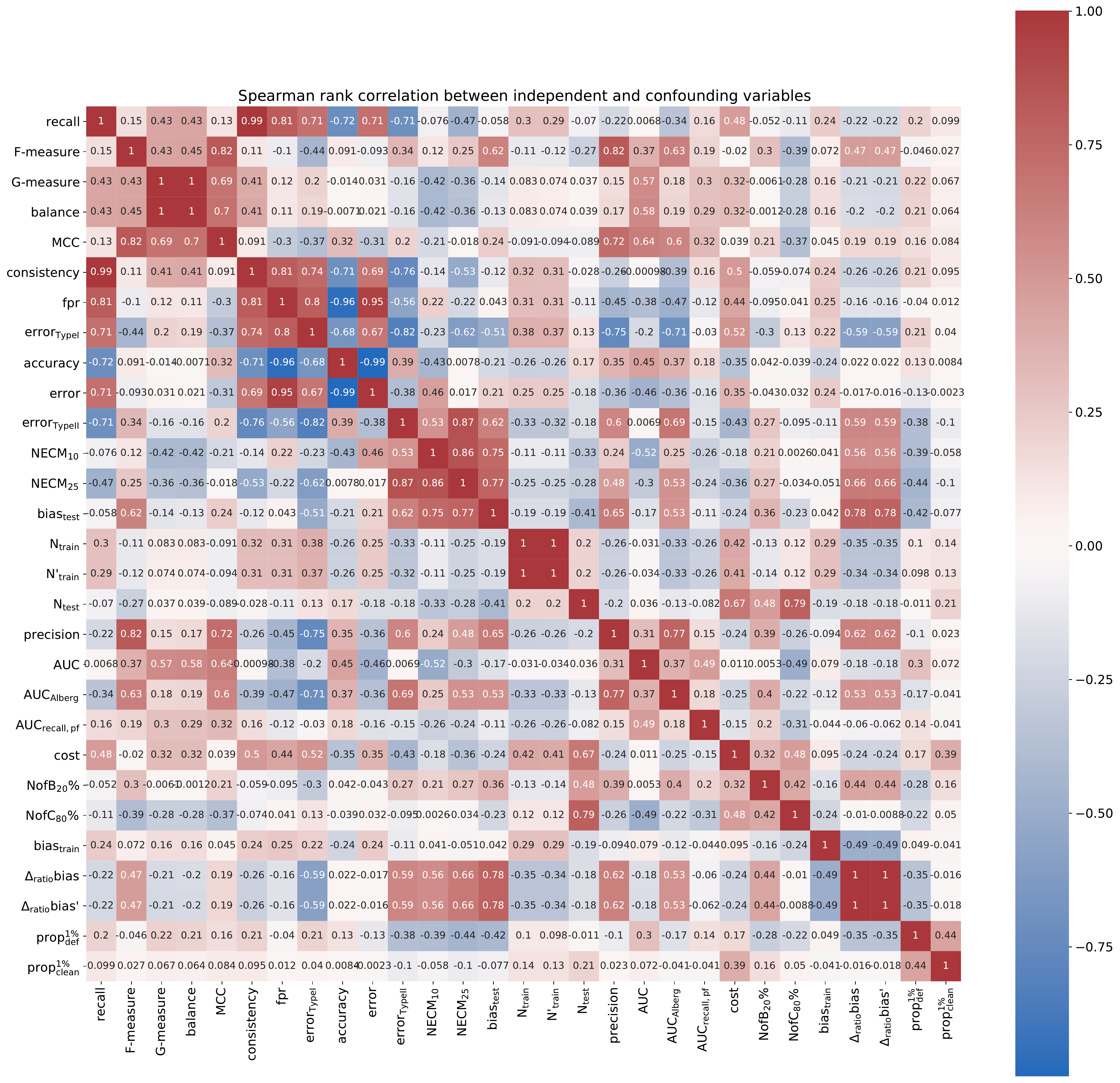}
\caption{Correlations between the variables (independent and confounding) in the generalization experiment. The variable order is the same as in Figure~\ref{fig:phase1_confmat} to enable a direct comparison.}
\label{fig:phase2_correlations}
\end{figure}

Figure~\ref{fig:phase2_correlations} shows the correlations between the independent and confounding variables in the generalization experiment. We observe that the correlation between the variables are not the same as in the bootstrap experiment. The notable differences are the following:
\begin{itemize}
    \item The \textit{recall} is still strongly correlated with \textit{consistency}. However, the other correlations are weaker now, especially to \textit{F-measure} and \textit{MCC}. Instead, the \textit{recall} now seems to be associated with \textit{fpr} $\textit{error}_{TypeI}$, \textit{accuracy}, \textit{error} and $\textit{error}_\textit{TypeII}$.
    \item The \textit{F-measure} is now correlated with the \textit{precision} instead of \textit{recall}.
    \item The \textit{accuracy} was correlated with the $\textit{error}_\textit{TypeII}$ in the bootstrap experiment. Now it is associated with the $\textit{error}_\textit{TypeI}$ as well. The correlation between the $\textit{bias}_\textit{test}$ is now low. 
\end{itemize}

\begin{figure}
\centering
\includegraphics[width=\textwidth]{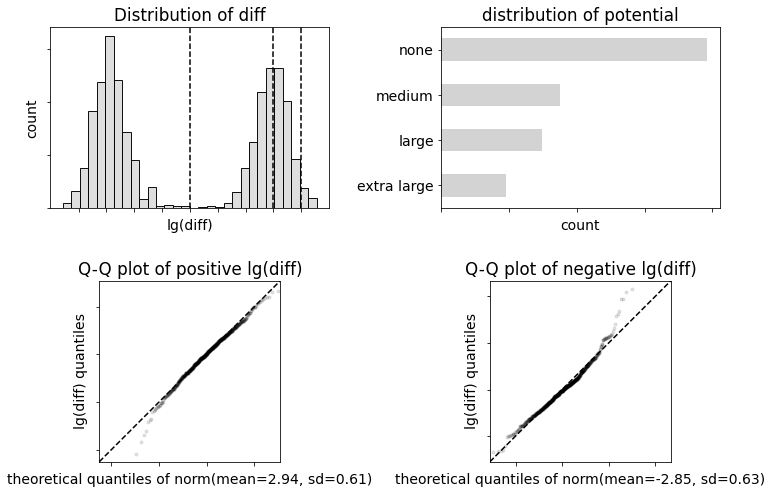}
\caption{Distribution of \textit{diff} and \textit{potential} in the cross-project and cross-version experiments. The Q-Q-plots are for normal distributions fitted to all values with positive, resp. negative values of \textit{diff}. The plots of \textit{diff} ignore 167 values of $+\infty$, 69 values of $-\infty$, and 5 \textit{NaN}s.}
\label{fig:phase2_distributions}
\end{figure}

Figure~\ref{fig:phase2_distributions} shows the distribution of the dependent variable in the generalization experiment, i.e., for the cross-version and cross-project predictions. The distribution of the positive values of \textit{diff} is similar, but with a slightly lower mean value and a slightly larger standard deviation. This leads to a more even distribution of the \textit{potential} between the classes \textit{medium}, \textit{large}, and \textit{extra large}. The distribution of the \textit{diff} that leads to a \textit{potential} of \textit{none} is completely different. In the bootstrap experiment, this was mostly driven by \textit{NaN}s, due not predicting any bugs. Now, the negative values of \textit{diff} are more or less symmetric to the positive values, i.e., the absolute values of \textit{diff} are similar, only the sign is different. We note that the distributions of \textit{diff} are similar for the cross-project and cross-version experiments, i.e., the combined data is representative.\footnote{The replication kit provides distribution plots separately for the cross-version and cross-project experiments. The distributions are the same, but the cross-version experiments have slightly more positive values of \textit{diff}.}

\begin{figure}
\centering
\includegraphics[width=\textwidth]{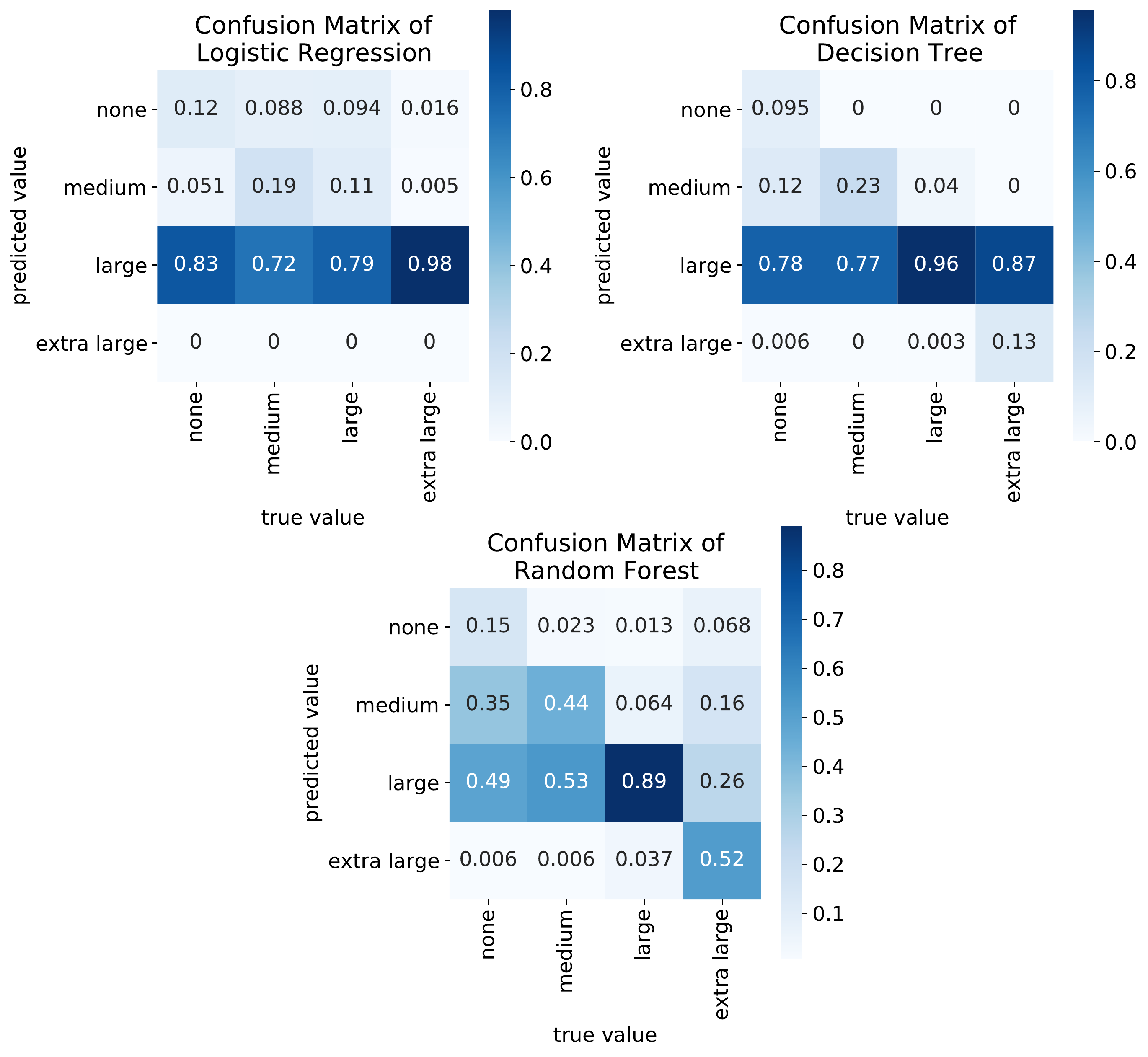}
\caption{Confusion matrices of the models trained on the bootstrap experiment with the cross-project and cross-version experiments as test data.}
\label{fig:phase2_confmat}
\end{figure}

\begin{figure}
\centering
\includegraphics[width=\textwidth]{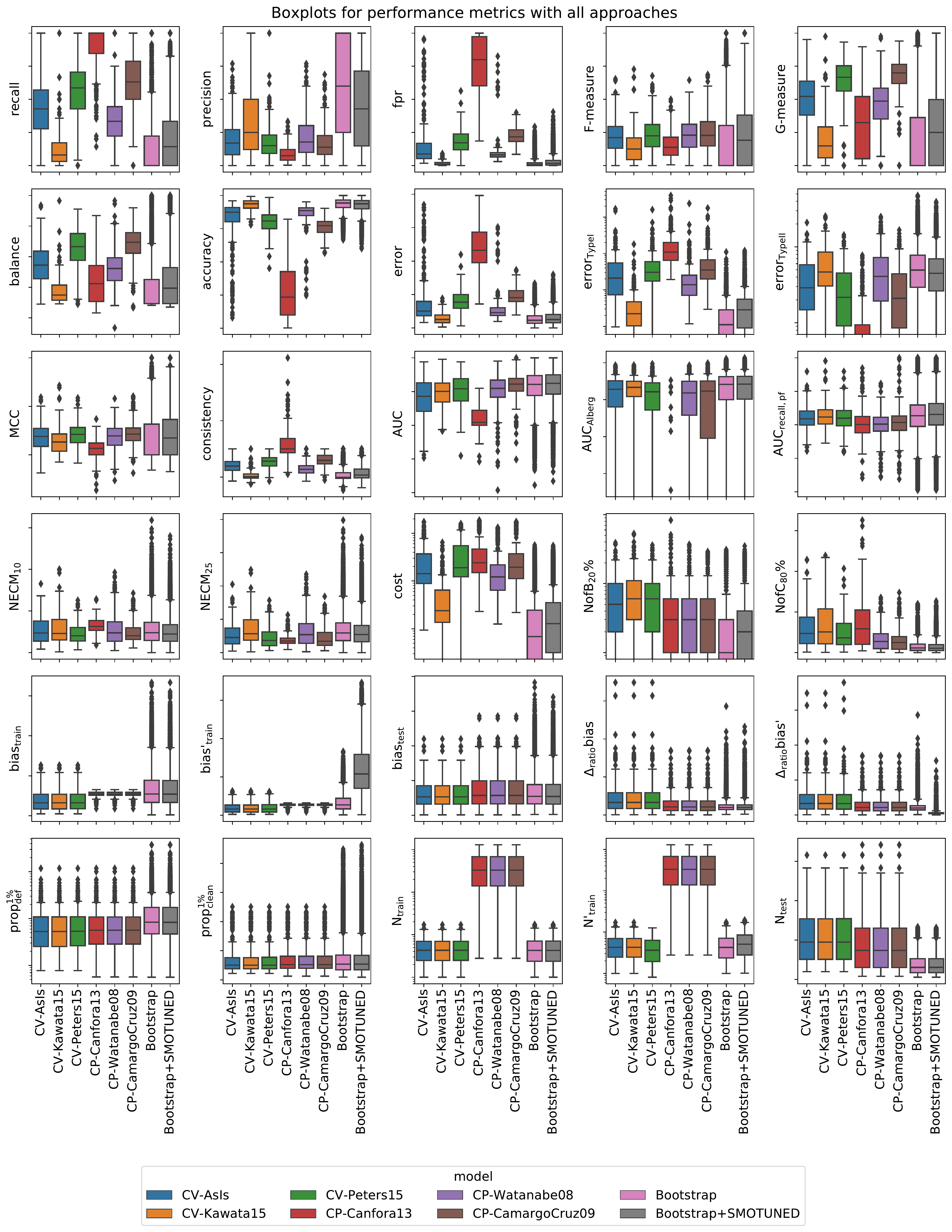}
\caption{Boxplots of the values of the independent and confounding variables. }
\label{fig:phase2_boxplot}
\end{figure}

Figure~\ref{fig:phase2_confmat} shows the results of applying the multinomial logit model, decision tree, and random forest trained as part of the bootstrap experiment to the cross-version and cross-project experiments. None of the models is good at the predictions of the class \textit{none}. The likely reason for this is that the models from the bootstrap experiment relied on very low values of variables from the \textit{recall} correlation group, that result from not predicting any instance as defective. Both the correlation analysis shown in Figure~\ref{fig:phase2_correlations} and the distribution of \textit{diff} shown in Figure~\ref{fig:phase2_distributions} indicate that this rarely happens within the generalization experiment. This is confirmed by the boxplots for the values we observe for every independent and confounding variable in both phases of the experiment (Figure~\ref{fig:phase2_boxplot}).

The positive classes \textit{medium}, \textit{large}, and \textit{extra large} are not affected by this strong shift in distributions. The main drivers we identified in the bootstrap experiment where \textit{accuracy}, $\textit{bias}_\textit{test}$, and $\textit{NECM}_{25}$, all of which still have similar distributions, according to Figure~\ref{fig:phase2_boxplot}. Nevertheless, the predictions of these classes are only mediocre: there are too many predictions as \textit{large}. While there is a small difference in the distributions of the \textit{diff} and \textit{potential} in comparison to the bootstrap experiment, such that the mean value of diff is a bit lower, the shift does not explain the magnitude of the missclassifications, especially considering that there was only moderate over and underprediction in the bootstrap experiment. In comparison, while the random forest has overall the most correct predictions (diagonal of the confusion matrix), there is strong tendency towards strong underprediction of \textit{extra large} values as \textit{medium} (16\% of \textit{extra large} instances affected). At the same time, there is also a strong tendency of moderate overprediction of \textit{medium} as \textit{large} (53\% of \textit{medium} instances affected). If the shift in distribution of the finite positive \textit{diff} would be responsible for this change, we would expect that we observe a trend towards either overprediction or underprediction, and not both. Consequently, this indicates that the models trained as part of the bootstrap experiment do not generalize to cross-version or cross-project defect prediction.

\subsection{Sensitivity Analysis}
\label{sec:sensitivity-analysis}

Our analysis of the distribution of \textit{diff} (see Figure~\ref{fig:dependent_bootstrap} and Figure~\ref{fig:phase2_distributions}) shows the choice of exponential bins for \textit{potential} is reasonable. However, we also observe that the distribution of the classes \textit{medium}, \textit{large} and \textit{extra large} sensitive to this choice: the central tendency of the log-normal distribution that describes the finite positive values of \textit{diff} is very close to $10^3$, i.e., the boundary between \textit{medium} and \textit{large}. Regardless, this does not seem to affect the capabilities of the random forest. Regardless of whether we conduct an upshift or downshift, the overall performance is almost exactly the same on the data from the bootstrap experiment.\footnote{The detailed results are part of the replication kit} Thus, the choice of bins does not change the capability of the model to predict the bins.

\begin{figure}
\centering
\includegraphics[width=\textwidth]{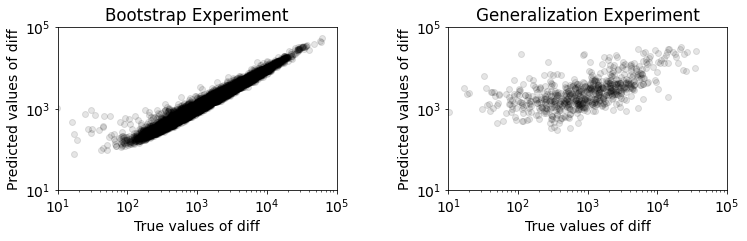}
\caption{Performance of a random forest regression model for \textit{diff} trained on the positive values of \textit{diff} from the bootstrap experiment. The left plot shows the performance for the bootstrap experiment, the right plot shows the performance on the generalization experiment.}
\label{fig:sensitivity_diff}
\end{figure}

To further validate that the lack of generalization from the bootstrap experiment to the cross-project experiment is not due our choice of bins, we trained a random forest regression model for \textit{diff} on the positive finite values of \textit{diff} from the bootstrap experiment. Figure~\ref{fig:sensitivity_diff} shows the result of this experiment. The random forest almost perfectly explains the bootstrap data, in line with the strong performance of the prediction model. However, the same random forest leads to almost random performance with the data from the generalization experiment, further demonstrating that the relation between the variables is different in both parts of the experiment, regardless of our design choice regarding the bins.

\section{Discussion}
\label{sec:discussion}

We now discuss the results of our study with respect to the relationship between the variables of our study, the mathematical explanation of the results, the defect prediction performance we observed, consequences for researchers, and threats to the validity of our work. 

\subsection{Relationship between Variables}
\label{sec:relationships}

Based on the criteria we established in Section~\ref{sec:theory} and the results we report regarding the bootstrap experiment in Section~\ref{sec:bootstrap}, we can state that there is a relationship between the variables that allows the modeling of the dependent variable on training data. While the multinomial logit model possibly was a bit aggressive in pruning variables and, as a result, was not able to defect the class \textit{extra large}, the decision tree and the random forest models were both able to yield good predictions within the bootstrap experiment. The performance is not quite at the level of strong categorization (90\% of instances in each class correct), but we easily fulfil the criteria for weak categorization within the bootstrap experiment. 

While this seems encouraging, all subsequent results rather point towards overfitting and a strong influence of the confounding variables, and not due to finding an actual relationship which can be used as proxy. The class \textit{none} is only identified by using a simple approach to find projects where nothing is predicted as defective. All variables used by the decision tree to distinguish between the classes \textit{medium}, \textit{large}, and \textit{extra large} are confounding variables, or strongly correlated to the confounding variable $\textit{bias}_\textit{test}$. The only result that generalizes from the bootstrap experiment to the generalization experiment is the distribution of \textit{diff}. Neither the prediction models, nor the correlations among the independent variables generalize. We note that the correlations among the confounding variables mostly generalize, except those directly measuring the data size, which was smaller for the test data in the bootstrap experiment (see Figure~\ref{fig:phase2_boxplot}). The sensitivity analysis confirmed that the lack of generalization is indeed due to a lack of generalization of the cost saving potential, and not due to our choice of an ordinal dependent variable. 

\begin{figure}
\centering
\includegraphics[width=\textwidth]{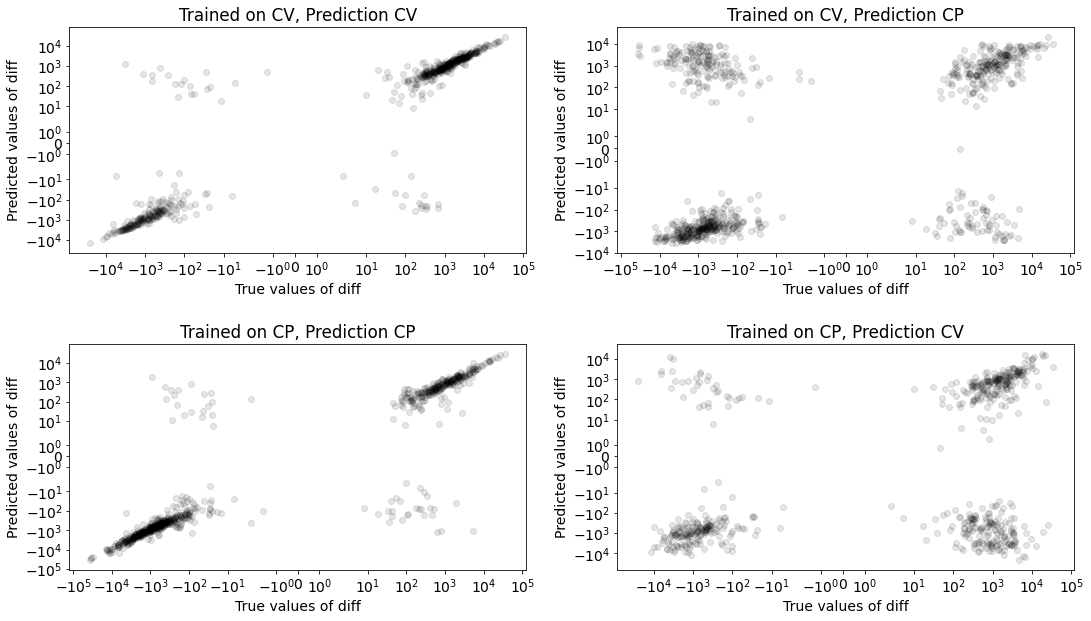}
\caption{Alternative random forest models trained on the data from the cross-version (CV) and the cross-project (CP) data.}
\label{fig:diff_models_cvcp}
\end{figure}

Due to this seeming lack of a strong relationship, other than overfitting to given results, we believe it is important to further establish that our protocol did not inadvertently cause the negative results, beyond the sensitivity analysis we already conducted. We identified one critical aspect, which could also explain why we did not find a generalizable relationship beyond overfitting. The bootstrap experiment was only suitable as a model for values of \textit{diff} larger than zero, i.e., positive cost potential. There were almost no negative values of \textit{diff}, instead the \textit{potential} of \textit{none} was rather due to trivial prediction models. Since this was different in the cross-version and cross-project experiments, it stands to reason, that the lack of generalization may be because the bootstrap experiment is not representative. In this case, we should be able to, e.g., find generalizable models when we train on the cross-version data and predict the cost saving potential of the cross-project data, or vice versa. However, as Figure~\ref{fig:diff_models_cvcp} indicates, this is not the case. Even when we use the relatively similar cross-version and cross-project predictions, using the same projects as test data, we observe the same: the random forest provides is a very accurate model on the training data, but this does not generalize. 

Due to all these results, we conclude the following regarding our research question: 

\begin{mdframed}
\textbf{RQ} Which performance metrics are good indicators for cost saving potential of defect prediction models?
\vspace{4pt}

\noindent
Performance metrics are not suitable indicators for the actual cost saving potential of defect prediction models. While it is possible to model the cost saving potential within an experiment, such models do not generalize and can, therefore, only be used in a limited way to establish relationships for a single experiment. Consequently, performance metric-based evaluations are, in general, neither suitable to determine if cost savings are possible, nor to determine the amount of cost saving potential. 
\end{mdframed}

\subsection{Mathematical Explanation of the Results}
\label{sec:math-explanation}

While our empirical data provides a clear indication for the result, there is no obvious reason why there is no generalization. We believe that reason is in the way costs are determined. Within this section, we provide a potential mathematical justification for this result. The variable \textit{diff} is defined as

\begin{align}
\textit{diff} &= \textit{upper}-\textit{lower} \\
&= \frac{\sum_{s \in S: h(s)=0} size(s)}{|D_{MISS}|} - \frac{\sum_{s \in S: h(s)=1} size(s)}{|D_{PRED}|}.
\end{align}

This formula is not directly comparable to the independent variables because it works with the sets $D_\textit{MISS}$ and $D_\textit{PRED}$, instead of the \textit{fn} and the \textit{tp}. If we simplify the costs and use the \textit{fn} and \textit{tp} instead, we have

\begin{equation}
\textit{diff}' = \frac{\sum_{s \in S: h(s)=0} size(s)}{fn} - \frac{\sum_{s \in S: h(s)=1} size(s)}{tp}.
\end{equation}

The second term, which is derived from the lower bound, is similar to the reciprocal \textit{precision}. This was already noted by \cite{Herbold2019} when the cost model was introduced. The difference is now essentially that the \textit{precision} uses $tp+fp$, which assumes that all instances have the same costs, while the cost model uses $\sum_{s \in S: h(s)=1} size(s)$, i.e., individual costs for each artifact, based on the size. A similar observation could be made for the first term, where the difference to confusion matrix-based approach is essentially that the cost model uses $\sum_{s \in S: h(s)=0} size(s)$ for individual costs instead of $tp+fn$. 

\begin{figure}
\centering
\includegraphics[width=0.5\textwidth]{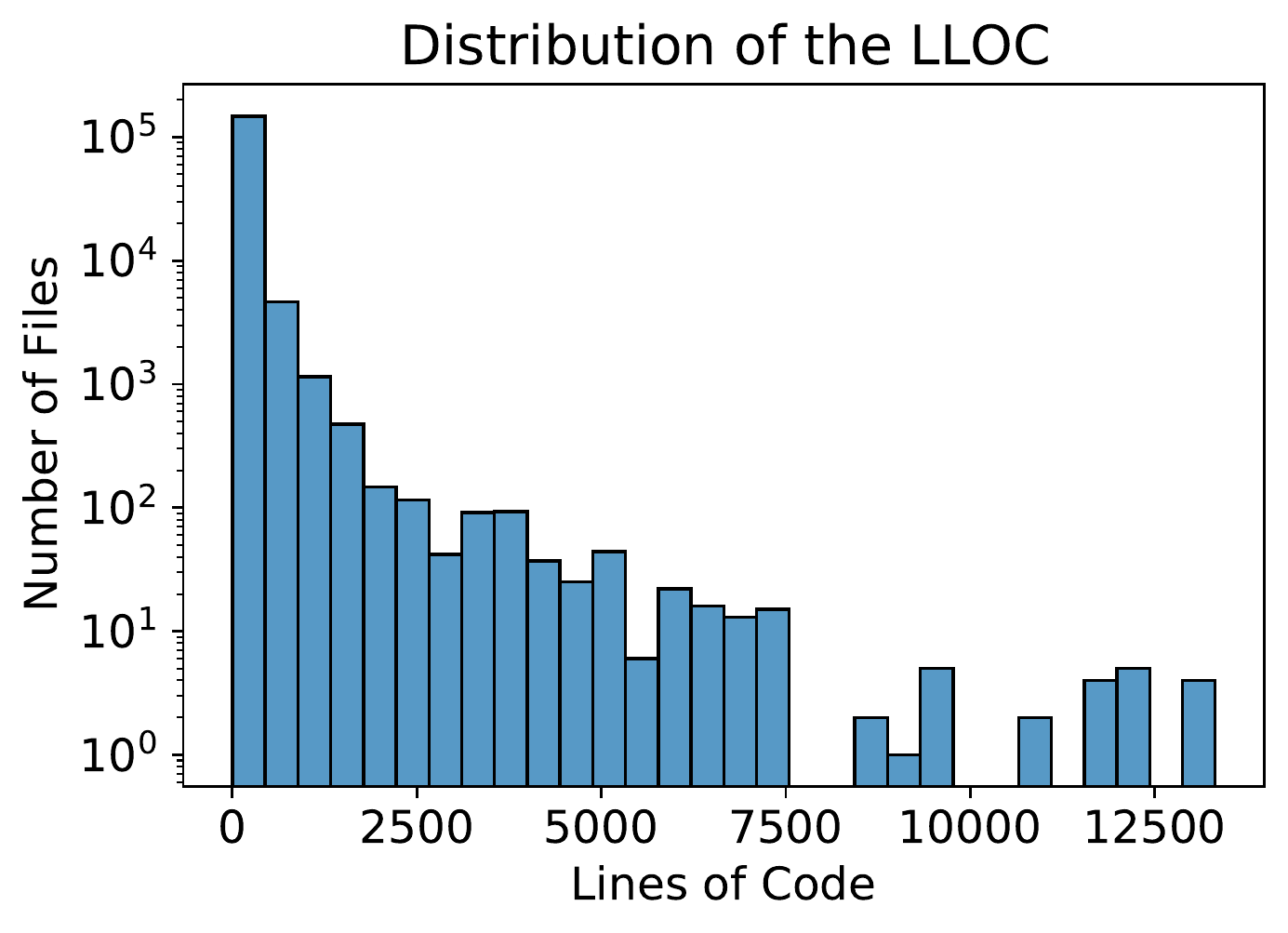}
\caption{Distribution of the Logical Lines of Code (LLOC) in the defect prediction data collected by \cite{Herbold2022}.}
\label{fig:lloc}
\end{figure}

Given this similarity, the usage of individual costs per artifact seems to make the difference: confusion matrices do not care which artifacts are predicted correctly, they only count them. However, predicting large artifacts correctly is, by the formula of the cost model, a lot more important, as they have a larger impact on the cases. Given that relatively few files have a large size (see Figure~\ref{fig:lloc}), this means that the few very large artifacts dominate the costs. The confounding variables $\textit{prop}_\textit{def}^{1\%}$ and $\textit{prop}_\textit{clean}^{1\%}$ were defined based on the idea to include the effect of very large files in our models. However, these variables only capture that such files are present, but not whether they are predicted correctly. 

Still, we also considered criteria that use the costs, most notably the variable $cost = \sum_{s \in S} h(s)\cdot size(s)$, which we can even rewrite as $cost = \sum_{s \in S: h(s)=1} size(s)$, which is the nominator of the second term of \textit{diff}, i.e., the lower bound. However, we have no similar proxy for the nominator of the first term. This is similar to only considering the \textit{precision}, but not the \textit{recall}, and therefore, allowing a trivial optimization. That such an incomplete consideration of the costs is no sufficient proxy for the total costs is, mathematically, not surprising. In this concrete case, this means that it is not sufficient to just consider the costs of quality assurance for the predictions, but that we also need to consider the costs we save by not applying quality assurance to the remainder of the project. 

In summary, we find similarity between the variables, but we can also identify the missing consideration of the individual costs for all artifacts as the likely reason for the lack of generalization. Thus, the mathematical analysis supports our conclusion from the empirical study that the independent and confounding variables are no suitable proxies for directly considering costs. 

\subsection{Defect Prediction Performance}

As a side effect, our study also provided the, to the best of our knowledge, first large-scale analysis of the cost saving potential of defect prediction models in realistic settings. The distribution of \textit{potential} and \textit{diff} we have seen earlier in Figure~\ref{fig:phase2_distributions} reveal two things:
\begin{itemize}
    \item Whether defect prediction can be cost saving at all is essentially a coin flip slightly favoring you, as the six approaches we used for the generalization experiment were not cost saving with a median of 46\%. The best result achieved 37\% (cross-version defect prediction the approach by \cite{Kawata2015}) and worst result achieved only 68\% (cross-project defect prediction with the approach by \cite{Watanabe2008}).
    \item If you win the coin flip (cost saving is possible), the range of actually cost saving values is log-normal distributed with a mean value of roughly 1000, which means that you require an estimation that is accurate to a single KLOC (kilo Lines of Code) of the relation between the costs for quality assurance and the cost of defects.
\end{itemize}

From our perspective, these results are not encouraging for practitioners willing to adopt defect prediction and show that feasible (release level) defect prediction that can actually save cost is still out of reach, at least when we apply models from the most recent benchmark studies we are aware of \citep{Amasaki2020, Herbold2017b}. While we cannot conclude that other models may not perform better, this first needs to be demonstrated with respect to costs in a study of similar size.

We note that these performance considerations are restricted to release-level defect prediction (predicting the defectiveness for all artifacts (e.g., classes, files) in a release and that these conclusions may not generalize to just-in-time defect prediction, i.e., the prediction if a change to a software introduced a new defect. The reason is that these results are regarding the quality of models, which may be different for just-in-time prediction. 

\subsection{Consequences for Defect Prediction Researchers}
\label{sec:recommendations}

As we discussed in sections~\ref{sec:relationships} and \ref{sec:math-explanation}, our study provides strong evidence that evaluations of defect prediction models must consider cost directly, if they should determine the quality of the models from an economic point of view. This does not mean, that costs should always be considered and that other metrics should not be considered anymore in future studies. In general, we believe there are two major kinds of different prediction studies: 

\begin{itemize}
    \item A new defect prediction model is proposed with the intent to demonstrate a better performance than the state of the art. Since a better prediction performance is directly related to the intent to be better from an economic point of view, such studies should use cost saving potential, or a similar criterion that directly measures costs, as main criterion for the comparison of approaches. Other metrics, e.g., \textit{recall}, may be used to augment such studies to provide insights into the behavior of the prediction model.
    \item A defect prediction model is used to study the relationship between a property (e.g., changes, static analysis warnings) and defects. This relationship is not only studied by pure prediction performance, but also by studying the inner workings of the defect prediction, to understand if and how the considered property is related to defects. Such studies do not need to consider the economic side of defect prediction and should instead follow the guidance by \cite{Yao2021} and use \textit{MCC}. 
\end{itemize}

We note that our experiments only consider release-level defect prediction. Thus, it is unclear if our conclusions generalize to just-in-time defect prediction. While it is possible that correlations between variables are more stable for just-in-time defect prediction, this opposite could also be true and our findings could generalize and just-in-time defect prediction could have equally unstable predictions. Moreover, even if the relationships between variables would be stable and there would be performance metrics as suitable proxies for cost savings, it would be unclear which metrics would be suitable and how the relationship would need to be interpreted. 

As researchers, we should be sceptical and not assume that using performance metrics is a valid approach from an economic point of view without evidence for this. Instead, we recommend to also follow our guidance established above regarding the two kinds of defect prediction studies for just-in-time defect prediction as well, unless a future study establishes performance metrics as suitable proxy for cost savings. This way, the threat to the validity of such studies due to unsuitable metrics for an economic assessment is mitigated. 

\subsection{Threats to Validity}

We report the threats to the validity of our work following the classification by \cite{Cook1979} suggested for software engineering by \cite{Wohlin2012}. Additionally, we discuss the reliability as suggested by \cite{Runeson2009}. 

\subsubsection{Construct Validity}

The construct of our study assumes that size is a suitable proxy for quality assurance effort, which may not be the case. In case this assumption does not hold, our results may be unreliable. \rev{While} we are not aware of any research that indicates that size is an unsuitable proxy for quality assurance effort\rev{, \cite{Shihab2013} found that complexity and combinations of metrics have stronger correlations to the effort for the correction of defects than the size. While this finding is not directly applicable to our work, because the cost model explicitly does not consider the costs for fixing a defect to be relevant for the cost effectiveness of defect prediction~\citep{Herbold2019}, this shows that the risk of using size for effort estimations is real and should not be underrated}. However, many metrics (e.g., complexity metrics) are correlated with the size~\citep{Mamun2019}. Thus, while size may not be a perfect proxy, the correlation of size with many metrics indicates that results are likely not invalid, even if there are better choices. As long as other factors (e.g., human factors such as experience) are correlated to size, the impact on our result should be similar to the impact of using different code metrics. However, if there would be no correlation to size, e.g., because different components of a larger system are affected differently, the cost modeling underlying our study would need to be revisited to understand the impact on the validity of our findings. Nevertheless, due to the current absence of such evidence we believe that the risk due to this assumption is an acceptable limitation to the construct of our study.

We mitigated further potential issues with our construct through a sensitivity analysis. Most notably, we determine if our choice of the levels of \textit{potential} impacts our results. We found that this was not the case. Moreover, we considered if the decision to use both bootstrap and more realistic settings affects our findings. However, we confirmed that we find similar, non-generalizable results when we fit models based on different data. 

Because there are many options in our case study design, e.g., the choice of defect prediction models, both for the bootstrap experiment, as well as for the cross-version and cross-project predictions, we pre-registered the study to get expert feedback on our construct and further mitigate the risk of issues with the validity of our construct. 

\subsubsection{Internal Validity}

One of our core conclusions is that the relationship between the variables do not generalize beyond the context of the training. We base this conclusion on the very good fit of models to the training data and the low performance on other data sets. A different explanation could be that this is not due differences between the context, but due to an approach to model fitting that favors overfitting and, thereby, hinders the generalization which would be possible with other models. However, we believe that this explanation is very unlikely because overfitting usually requires that models are too complex, which is certainly not the case for the three-variable multinomial logit model, but also not for the decision tree, where we can directly observe that most decisions in the tree affect large amounts of data and provide clear partitions. With overfitting, we would expect more decisions in the tree that only affect a small subset of the data. Moreover, random forests are in general relatively robust against overfitting, due to the ensemble trained on bootstrap samples (overfitting would be restricted to subsets) and subsets of variables (overfitting would need to be possible on subsets). Moreover, we support our conclusion not just through the empirical data, but also by outlining why the mathematical properties suggest that the independent and confounding variables are insufficient to model the cost saving potential. 

\subsubsection{External Validity}

We already discuss the generalizabilty to just-in-time defect prediction in Section~\ref{sec:recommendations}. Beyond that, it is also unclear without an independent confirmatory study if our results hold for other programming languages or even other samples of projects written in Java. However, we believe that our results should hold and may even be relevant beyond defect prediction research, as other prediction models may also be affected. Our mathematical analysis suggests that if different costs are associated not only with different classes, but also with each instance, they should be considered for an accurate economic assessment of models. In our case, the costs follow an exponential distribution, meaning that a relatively small proportion of data drives most of the costs. We believe that other scenarios with such unequal costs should use cost modeling for the evaluation, as other criteria provide an incomplete picture. For example, for the prediction of creditworthiness \citep{Huang2007}, the amount of money of a loan should be considered.

\subsubsection{Reliability}

Our exploratory study was conducted by two researchers, who brought in their own experience working with such data. Other researchers may look at different aspects, which could lead to different conclusions. To mitigate the impact of this, the study protocol was pre-registered with the goal to ensure that the approach by the researchers does not bias the results towards certain findings. 

\section{Conclusion}
\label{sec:conclusion}

Within this article, we considered the question if the performance metrics that are usually used in defect prediction research are suitable for the evaluation of defect prediction model in terms of costs, i.e., from an economic point of view. We conducted several experiments and found that we could not establish a relationship between the used criteria and costs. A mathematical analysis of the costs and the performance metrics reveals that the exponential distribution of the size of software artifacts is the likely reason for this lack of a relationship because this means that the correct prediction of relatively few artifacts drives most of the costs. Our results mean that costs should always be considered directly, and not through proxy metrics. Moreover, we found that the defect prediction models we studied have a large risk of at least 37\% not being cost saving. 

\rev{Future work should attempt to fill our gaps in understanding the costs and benefits of defect prediction and related maintenance activities. Notably, a better validation of the underlying costs, especially regarding the usage of size as a proxy for costs for quality assurance, as well as a more accurate assessment of the underlying costs of defects could aid defect prediction researchers to better understand the requirements that defect prediction models need to fulfill to be cost efficient. Moreover, our considerations are strictly focused on costs of quality assurance effort. Notably, our cost modeling assumes that a false positive only consumes money without a benefit. In reality, there could be benefits, they would just not be related to the defect prediction. For example, while the effort invested in quality assurance for false positive predictions would not lead to the identification of defects, this could still lead to a better test coverage or less technical debt within the source code. Similarly, developers could learn something about the code even when no defect is discovered. This could lead to lower maintenance costs in the future. However, such aspects are hard to quantify. Longitudinal studies of real projects with interventions through defect prediction could shed light on such topics, to determine if false positives are really a problem while also better understanding the impact of quality assurance.}

\rev{Moreover, researchers that want to study the efficiency of defect prediction should go beyond the reporting of model quality and also consider the possible interventions, i.e., how developers should and could react on prediction. Past research showed that this is crucial for the adoption by developers~\citep[e.g.][]{Lewis2013} and recent studies on explanations of predictions~\citep[e.g.][]{Jiarpakdee2022} provide an interesting future venue that also shows potential benefits of models for risk assessment confusion-matrix style evaluations.}

\bibliography{./literature}

\end{document}